\documentclass[sigplan,10pt,screen,review=false,anonymous=false]{acmart}
\AtBeginDocument{%
  }

\usepackage{acmart-taps}
\usepackage{enumitem}
\usepackage{array}
\usepackage{graphicx}
\usepackage{subcaption}
\usepackage{booktabs}
\usepackage{makecell}
\usepackage{pifont}
\usepackage{multirow}
\usepackage[utf8]{inputenc}
\usepackage{algorithm}
\usepackage{algorithmic}
\usepackage{amsmath}
\usepackage{booktabs}
\usepackage{multirow}

\usepackage{pifont}          
\usepackage{algorithm}
\usepackage{booktabs}        
\usepackage{amsmath}

\newcommand{\chen}[1]{\textcolor{red}{#1}}
\renewcommand{\chen}[1]{#1}
\newcommand{\nchen}[1]{#1}

\begin{document}


\copyrightyear{2026}\acmYear{2026}
\setcopyright{cc}
\setcctype{by}
\acmConference[SIGOPS ATC 2026]{The 2026 ACM SIGOPS Annual Technical Conference}{November 15-18, 2026}{Hong Kong, China}
\acmBooktitle{The 2026 ACM SIGOPS Annual Technical Conference (SIGOPS ATC 2026), November 15-18, 2026, Hong Kong, China}
\acmDOI{10.1145/3828523.3857067}
\acmISBN{979-8-4007-2818-1/2026/11}

\title{InfiniLoRA: Disaggregated Multi-LoRA Serving for Large Language Models}

\author{Hongyu Chen}
\email{chenhongyu2048@sjtu.edu.cn}
\affiliation{%
  \institution{Shanghai Jiao Tong University}
  \city{Shanghai}
  \country{China}
}
\author{Letian Ruan}
\email{1291903308rlt@sjtu.edu.cn}
\affiliation{%
  \institution{Shanghai Jiao Tong University}
  \city{Shanghai}
  \country{China}
}
\author{Zilin Xu}
\email{xuzilin.200415@bytedance.com}
\affiliation{%
  \institution{ByteDance}
  \city{Beijing}
  \country{China}
}
\author{Yuchen Li}
\email{yuchenli@smu.edu.sg}
\affiliation{%
  \institution{Singapore Management University}
  \country{Singapore}
}
\author{Xinyu Chen}
\email{xinyuchen@hkust-gz.edu.cn}
\affiliation{%
  \institution{The Hong Kong University of Science and Technology (Guangzhou)}
  \city{Guangzhou}
  \country{China}
}
\author{Bingsheng He}
\email{dcsheb@nus.edu.sg}
\affiliation{%
  \institution{National University of Singapore}
  \country{Singapore}
}
\author{Jingwen Leng}
\email{leng-jw@sjtu.edu.cn}
\affiliation{%
  \institution{Shanghai Jiao Tong University}
  \city{Shanghai}
  \country{China}
}
\author{Jieru Zhao}
\email{zhao-jieru@sjtu.edu.cn}
\affiliation{%
  \institution{Shanghai Jiao Tong University}
  \city{Shanghai}
  \country{China}
}
\author{Shixuan Sun}
\authornote{Shixuan Sun is the corresponding author.}
\email{sunshixuan@sjtu.edu.cn}
\affiliation{%
  \institution{Shanghai Jiao Tong University}
  \city{Shanghai}
  \country{China}
}
\author{Minyi Guo}
\email{guo-my@cs.sjtu.edu.cn}
\affiliation{%
  \institution{Shanghai Jiao Tong University}
  \city{Shanghai}
  \country{China}
}

\renewcommand{\shortauthors}{Hongyu Chen et al.}

\begin{abstract}
    LoRA enables efficient customization of LLMs and is widely used in multi-tenant and multi-task serving. However, emerging model architectures such as MoE significantly increase LoRA memory cost, making existing coupled LoRA serving designs poorly scalable and prone to tail-latency inflation. We present InfiniLoRA, a disaggregated LoRA serving system that decouples LoRA execution from base-model inference. InfiniLoRA introduces a shared LoRA Server with parallelism-aware execution, SLO-driven provisioning, and critical-path optimizations, including GPU-initiated communication and hardware-specialized LoRA kernels. Experiments show that InfiniLoRA can achieve an average $3.05\times$ increase in serviceable request rate under strict latency SLOs, and improve the percentage of LoRA adapters satisfying the SLO requirement by 54.0\%.
\end{abstract}



\newlength{\savedemergencystretch}
\setlength{\savedemergencystretch}{\emergencystretch}
\setlength{\emergencystretch}{1em}
\maketitle
\setlength{\emergencystretch}{\savedemergencystretch}
\pagestyle{empty}
\thispagestyle{empty}

\section{Introduction} \label{sec:introduction}


Low-Rank Adaptation (LoRA)~\cite{hu2022lora, Dettmers2023QLoRA, wang2024lorapro, shen2023mixtureofexpertsmeetsinstructiontuninga} has become an important building block for deploying large language models (LLMs) in real-world systems. By enabling parameter-efficient fine-tuning, LoRA allows LLMs to incrementally incorporate task-specific, domain-specific, or user-specific knowledge without retraining or replicating the full model. This capability is particularly important for stateful applications, such as long-term memory~\cite{anonymous2026understandinglora, wang2025mextendingmemoryllmscalable, chen2024longloraefficientfinetuninglongcontext}, personalization~\cite{li2024personalllmagentsinsights, li2025helloagainllmpoweredpersonalized, Zhang2024PersonalizedLoRA}, and preference modeling~\cite{chen2025mapreduceloraadvancingpareto, Kong2024iloraCustomizing, yang2025loraliteecomputationallyefficientframework}, where models are expected to retain, update, and apply behavioral or contextual information across interactions.

In cloud-based LLM serving platforms, these properties make LoRA well suited for multi-tenant scenarios. A single base model can be shared across many tenants or applications, while different LoRA adapters are dynamically activated to encode memories, skills, roles, or domain expertise on demand~\cite{chen2023punicamultitenantloraserving}. The serving frameworks further batch requests targeting different adapters to amortize computation and improve throughput~\cite{sheng2023slora, dLoRA24osdi}.

\begin{figure}[tbp]
    \vspace{-4pt}
    \centering
    \begin{subfigure}{0.95\linewidth}
        \centering
        \includegraphics[width=\linewidth]{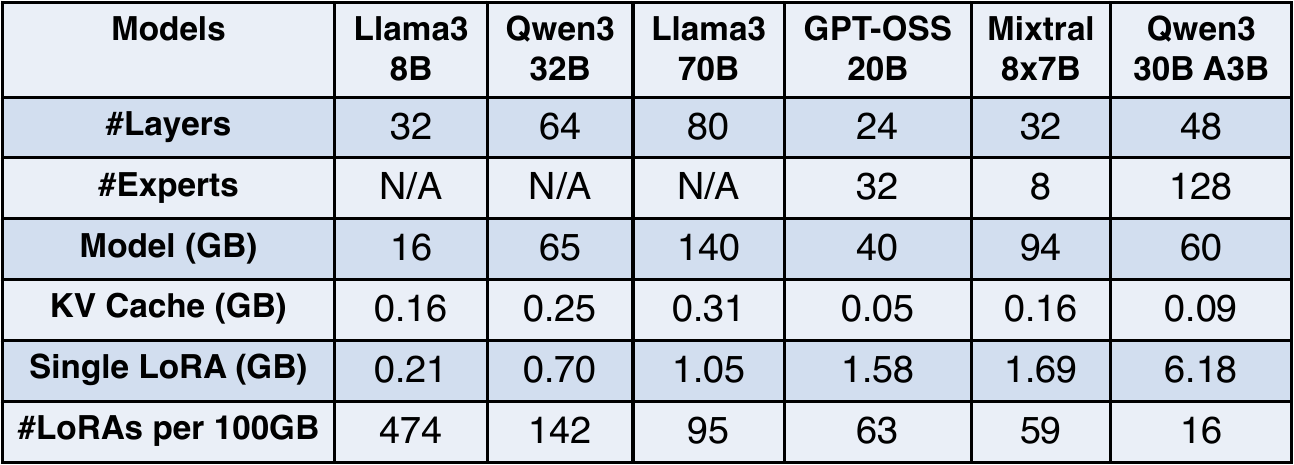}
        \vspace{-13pt}
        \caption{Memory consumption of model weights, KV cache (1024 tokens), and LoRA for representative dense and MoE models, with LoRA rank=64.}
        \label{fig:dense_moe_capacity_comparison}
    \end{subfigure}
    \begin{subfigure}{0.95\linewidth}
        \centering
        \includegraphics[width=\linewidth]{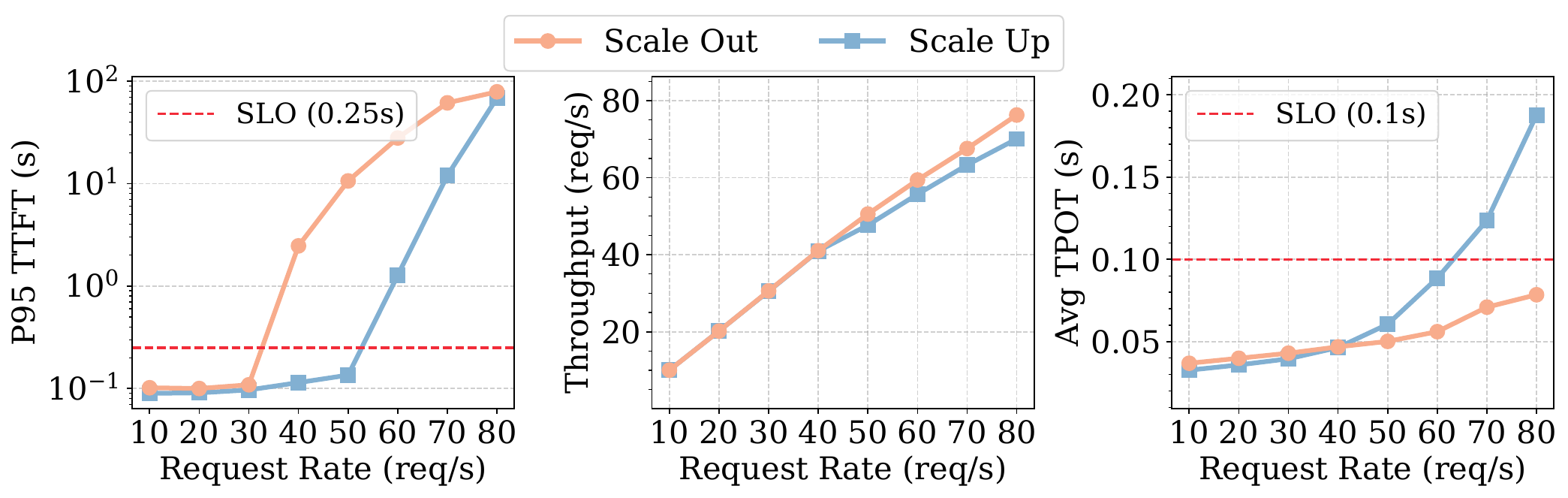}
        \vspace{-13pt}
        \caption{Scale-out (Eight 2-GPU Instances) vs. Scale-up (Four 4-GPU Instances) performance comparison under varying load on Nvidia Hopper GPUs with Mixtral~8x7B model and 512 LoRAs. The red dashed line indicates SLO.}
        \label{fig:motivation_scale_up_out}
    \end{subfigure}
        \vspace{-10pt}
    \caption{(\textbf{Top}) LoRA cache capacity across model architectures. (\textbf{Bottom}) Scale-out vs. scale-up performance.}
    \label{fig:arch_evolution}
    \vspace{-6pt}
\end{figure}

As LLM architectures continue to evolve, however, the assumptions underlying existing LoRA serving designs are increasingly challenged. Emerging architectures, most notably Mixture-of-Experts (MoE) models~\cite{fedus2022switchtransformersscalingtrillion, jiang2024mixtralexperts, deepseekai2025deepseekv3technicalreport, Hu2025MoEFinetuning, shi2025expertweaveefficientlyservingexpertspecialized, shen2025EdgeLoRA}, significantly increase the parameter footprint of LoRA adapters by introducing expert-specific adaptations, as illustrated in Figure~\ref{fig:dense_moe_capacity_comparison}. Meanwhile, modern workloads demand longer context windows and richer interaction histories, substantially enlarging the KV cache footprint~\cite{shi2024incontext, xiao2024duoattentionefficientlongcontextllm, gao2025prolong}. The combined pressure from adapter expansion and KV cache growth severely constrains GPU memory resources, fundamentally limiting the capacity available for serving LoRA adapters.
    
\noindent\textbf{Limitations of Existing LoRA Serving Designs.}
Existing LoRA serving frameworks~\cite{chen2023punicamultitenantloraserving, sheng2023slora, dLoRA24osdi, Iliakopoulou2025Chameleon, li2025toppings, zhang2025improvingservingperformancemultilora, Xia2024Efficientmultitask} adopt a \emph{coupled design}, where LoRA adapters are co-located in GPU memory with the base model weights and KV cache to avoid adapter loading overheads. 
\chen{Such coupled design and local execution is attractive when the active adapter working set--the set of distinct adapters needed by concurrent requests--fits in the local cache. Larger expert-specific MoE adapters reach this capacity limit sooner because of their greater memory footprint (Figure~\ref{fig:dense_moe_capacity_comparison}).}
Consequently, requests targeting non-resident adapters must wait for memory to be reclaimed before their adapters can be loaded, introducing substantial queueing delays. Since Time-to-First-Token (TTFT) is highly sensitive to such delays, this leads to a significant degradation in tail latency.

A natural response is to increase LoRA cache capacity by either \emph{scaling out}, i.e., deploying more LLM instances, or \emph{scaling up}, i.e., allocating more GPUs to a single instance. However, both approaches have fundamental limitations. Scaling out increases total cache capacity across instances but requires duplicating base model weights for each instance, incurring substantial GPU memory overhead. Given the size of modern LLMs, this duplication leaves little additional memory for LoRA, resulting in only marginal cache gains. As shown in Figure~\ref{fig:motivation_scale_up_out}, P95 TTFT quickly degrades and violates the TTFT SLO even when using eight 2-GPU instances, due to limited effective cache capacity.

Scaling up aggregates more GPUs within a single instance, expanding the LoRA cache without duplicating base weights. However, it enlarges the communication scope: as the instance spans more GPUs, especially across nodes, communication overheads grow rapidly and inflate inference latency. Moreover, efficiently utilizing the increased resources requires larger batch sizes, which are often incompatible with latency-sensitive workloads. Consequently, as shown in Figure~\ref{fig:motivation_scale_up_out}, scaling up achieves lower throughput than scaling out and suffers significantly higher average Time-Per-Output-Token (TPOT). In summary, the core limitation of existing approaches stems from a \textit{coupled} design that tightly binds LoRA adapters to the base model within the individual LLM instances. As a result, accommodating dynamic LoRA workloads requires modifying base-model execution, which limits flexibility and leads to inefficient resource utilization.

\noindent\textbf{Our Approach.}
We propose \textbf{InfiniLoRA}, a \emph{disaggregated LoRA serving architecture} that decouples LoRA adapters from LLM instances. Adapters are managed by a dedicated LoRA Server, while LLM instances remain LoRA-free and focus on base-model inference. This decoupling enables independent scaling of LoRA resources, but introduces  a key implication: remote LoRA execution incurs additional communication overhead. 
\chen{Our key insight is that this overhead can be largely hidden via \emph{dual-level pipelining}: at the LLM side, LoRA communication and computation are overlapped with MoE execution; at the server side, communication from one instance is overlapped with computation for others.} 
This transforms remote LoRA execution from a latency overhead into an opportunity for communication-computation overlapping. Realizing this efficiently requires addressing three tightly coupled challenges.

\emph{Challenge 1: Parallelism and Placement for Pipeline Efficiency.}
Disaggregation introduces an M-to-N interaction between LLM instances and the LoRA Server. LoRA data placement and parallelism directly determine communication volume, computation parallelism, and synchronization scope, thereby governing the achievable overlap.

\emph{Challenge 2: Critical-Path Optimization for Decode-Time Execution.}
LoRA execution remains on the decode-time critical path and directly affects TPOT. Additional communication and synchronization can degrade latency if not carefully optimized. The challenge is to integrate remote LoRA execution while preserving the benefits of overlap.

\emph{Challenge 3: SLO-Aware Provisioning for Sustained Pipeline Throughput.}
Pipeline efficiency depends on sufficient LoRA Server capacity under dynamic workloads. Under-provisioning causes queueing and breaks overlap, while over-provisioning wastes resources. The system must provision resources in an SLO-aware manner to sustain throughput and meet TTFT and TPOT targets.

To address these challenges, InfiniLoRA adopts a unified design combining parallelism-aware execution, critical-path optimization, and SLO-driven provisioning. It employs a hybrid execution strategy that integrates expert parallelism with pipeline parallelism to enable efficient dual-level pipelining while balancing communication, synchronization, and GPU utilization. It further performs SLO-driven provisioning to determine minimal cache capacity and GPU resources, and optimizes the decode-time path via GPU-initiated communication, specialized LoRA kernels, and pipelined adapter loading with prefetching. Using both production traces and synthetic workloads, InfiniLoRA significantly outperforms state-of-the-art multi-LoRA serving systems, sustaining up to $3.05\times$ higher serviceable request rates under strict TTFT and TPOT SLOs and improving the fraction of adapters meeting stringent SLO targets (over 90\%) by 54.0\%. In summary, this paper makes the following contributions:

\begin{itemize}[leftmargin=*]
    \item
    We identify the scalability limitation of coupled LoRA serving and propose InfiniLoRA,  decoupling LoRA adapters from LLM instances to enable independent scaling.

    \item
    We introduce a pipelined execution model that overlaps LoRA communication and computation across LLM instances and the LoRA Server, turning remote execution into pipeline parallelism.
    
    \item
    We present a unified design for disaggregated LoRA serving, including parallelism-aware execution, critical-path optimization, and SLO-driven provisioning.

    \item InfiniLoRA achieves up to $3.05\times$ higher serviceable request rates and improves SLO attainment by 54.0\% over state-of-the-art systems.
\end{itemize}
\section{Background}
\subsection{LLM Inference}
\label{sec:background_llm_inference}

\begin{figure}[tbp]
    \vspace{-4pt}
    \centering
    \includegraphics[width=0.95\columnwidth]{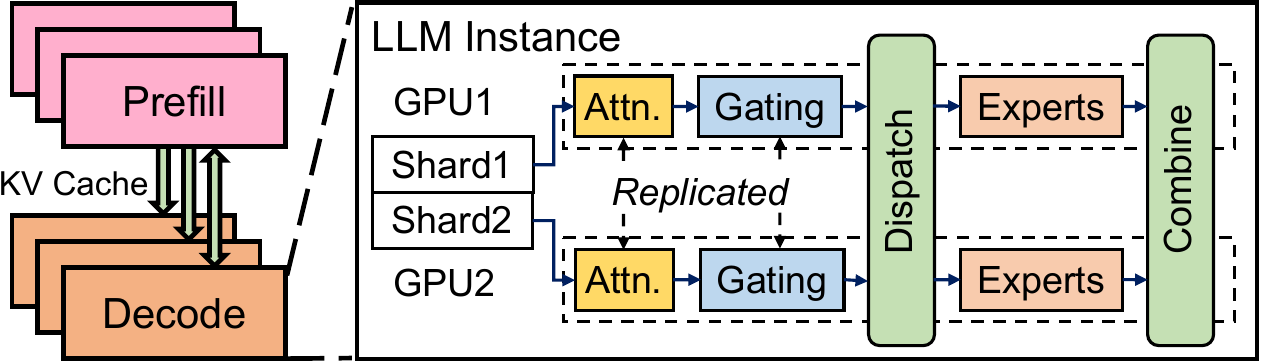}
        \vspace{-10pt}
    \caption{Prefill–decode disaggregated architecture. LLM instances are deployed with 2 GPUs using expert parallelism.}
    \Description{Need to modify, just a simplified diagram}
    \label{fig:pd_disaggregation}
    \vspace{-13pt}
\end{figure}

LLM inference consists of two phases: \emph{prefill} and \emph{decode}. The prefill phase processes the entire input prompt in parallel to initialize the KV cache and is compute-bound due to dense matrix multiplications. The decode phase generates tokens autoregressively, repeatedly accessing the KV cache, making it memory-bandwidth-bound with limited per-request parallelism. To avoid interference between the two phases, recent serving systems~\cite{Pratyush2024splitwise, Zhong2024distserve, Hu2025ShuffleInfer, qin2025mooncake, hu2024memservecontextcachingdisaggregated} adopt a prefill–decode disaggregated architecture (Figure~\ref{fig:pd_disaggregation}), assigning them to separate resources. In practice, a prefill GPU achieves high utilization even with a single request due to its high arithmetic intensity, whereas decode GPUs require batching tens to hundreds of concurrent requests to saturate memory bandwidth.

During decoding, large models employ hybrid parallelism~\cite{rajbhandari2022deepspeedmoeadvancingmixtureofexpertsinference, zhang2025bestpracticesforsearvingdeepseekr1}. For MoE models, \emph{expert parallelism} partitions experts across GPUs, while attention layers use \emph{data parallelism}, sharding the input batch and routing tokens to the corresponding experts. \emph{Tensor parallelism} may further split attention computation across devices.

\subsection{LoRA Computation and Multi-LoRA Serving} \label{sec:lora_serving_background}

For a weight matrix $W \in \mathbb{R}^{h \times d}$, LoRA applies a low-rank update $W' = W + AB$, where $A \in \mathbb{R}^{h \times r}$ and $B \in \mathbb{R}^{r \times d}$ are trainable \emph{adapter} matrices. The output for input $x$ becomes $y' = xW + xAB$. The rank $r$ is typically small (e.g., 32--128), greatly reducing both training and inference cost compared to full fine-tuning~\cite{hu2022lora, Dettmers2023QLoRA, schulman2025lora}.

\begin{figure}[tbp]
    \vspace{-4pt}
    \centering
    \includegraphics[width=0.95\columnwidth]{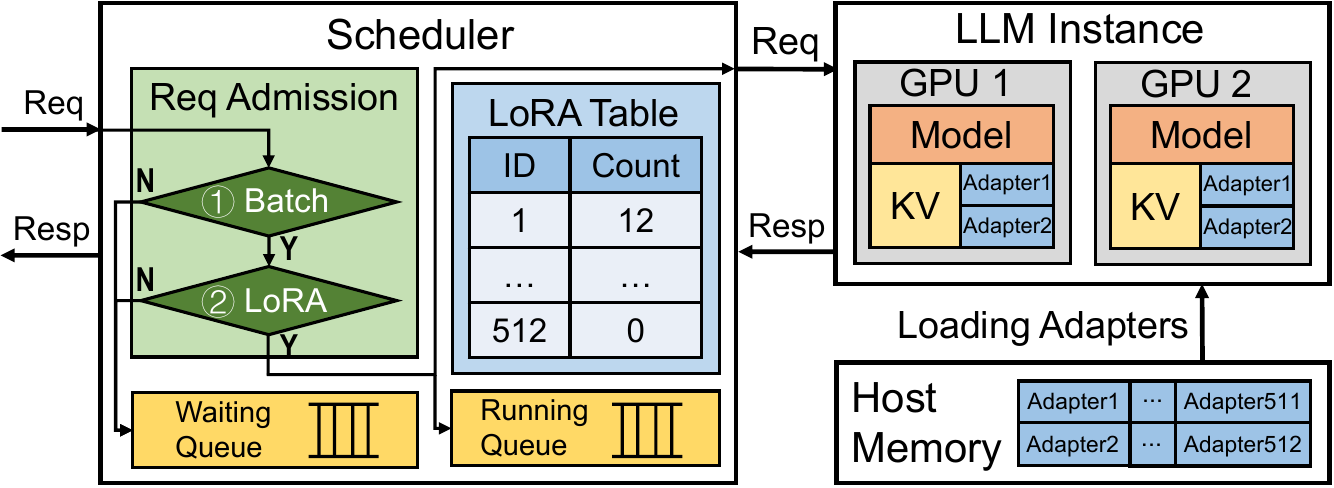}
        \vspace{-10pt}
    \caption{Coupled-design multi-LoRA serving architecture.}
    \Description{Need to modify, just a simplified diagram}
    \label{fig:multi_lora_serving}
    \vspace{-10pt}
\end{figure}

In multi-task and multi-tenant environments, many adapters are served concurrently, with different requests in the same batch potentially requiring different adapters. Recent systems~\cite{chen2023punicamultitenantloraserving, sheng2023slora, dLoRA24osdi, Iliakopoulou2025Chameleon, li2025toppings, zhang2025improvingservingperformancemultilora, kwon2023vllm} exploit the shared base model by consolidating such requests into a single batch: the base computation ($xW$) is batched, while each request independently computes its LoRA update ($xAB$) and adds it to the base output. Existing systems adopt a \emph{coupled architecture} in which adapters reside in GPU memory alongside base weights and KV cache. As shown in Figure~\ref{fig:multi_lora_serving}, a LoRA-aware scheduler manages request admission.

Upon receiving a request, the scheduler first checks whether the engine has reached its batch capacity, bounded by KV cache limits and TPOT SLOs. If admissible, it consults the LoRA table and admits the request only if the required adapter is resident or can be loaded within available cache space; otherwise, the request is queued. Scheduling operates at token granularity, enabling per-step admission and retirement of requests while dynamically updating the LoRA table.

Under this coupled architecture, prior work improves performance through several directions: optimizing cache replacement policies to increase LoRA cache hit rates~\cite{Iliakopoulou2025Chameleon, zhu2025Cannikin, Wu2025rockserving}, redesigning data layouts to reduce adapter loading overhead~\cite{li2025toppings, Shi2025aulora}, rebalancing GPU memory allocation between KV cache and LoRA cache~\cite{zhang2025improvingservingperformancemultilora}, and developing specialized GPU kernels to accelerate LoRA computation when requests within a batch require different adapters that scattered across memory~\cite{Zhou_Zhou_Zhang_Wang_Liu_2025, Xia2024AdaptiveOperatorScheduling, kong2024loraswitchboostingefficiencydynamic}.

LoRA serving during prefill is relatively straightforward: the batch size is typically small (e.g., 1–4 requests)~\cite{Zhong2024distserve, du2025prefillonly}, limiting LoRA cache overhead, and known input lengths allow adapters to be prefetched by overlapping loading with prior-batch computation. In contrast, decode-phase LoRA serving is substantially harder due to larger batches and unpredictable decoding lengths. Consequently, both prior work and our design focus primarily on optimizing LoRA serving in the decode phase.

\subsection{MoE-Induced Issues in Multi-LoRA Serving}
\label{sec:motivation_issues}

As LLMs evolve, the MoE architecture has become increasingly prevalent and is adopted by recent models such as Mixtral~\cite{jiang2024mixtralexperts}, Qwen3~\cite{qwen3technicalreport}, and DeepSeek~\cite{deepseekai2025deepseekv3technicalreport}. In MoE models, LoRA adapters maintain expert-specific parameters, causing adapter size to scale with the number of experts. As the expert count grows, the memory footprint of each adapter increases substantially. Consequently, under a fixed LoRA memory budget, the number of adapters that can reside on GPUs drops sharply (Figure~\ref{fig:dense_moe_capacity_comparison}), significantly undermining the effectiveness of multi-LoRA serving.

To quantify this effect, we evaluate S-LoRA~\cite{sheng2023slora}, using its state-of-the-art implementation integrated into vLLM~\cite{kwon2023vllm}, as opposed to the original vanilla codebase, on the Mixtral 8x7B model. We deploy four LLM instances, each running on two NVIDIA Hopper GPUs (96 GB), and vary the LoRA cache ratio from 10\% to 50\%, where the cache ratio denotes the fraction of adapters that can be simultaneously resident in GPU memory across all instances. The system serves 256 LoRA adapters, each consuming 1.69 GB of GPU memory, with access frequencies following a Zipf distribution ($s=\text{1.2}$) as in prior work~\cite{sheng2023slora, chen2023punicamultitenantloraserving}. The P95 TTFT SLO is set to 0.25 s. This setup isolates the impact of limited LoRA cache capacity in coupled architectures. Our results reveal two key issues that arise when cache capacity is constrained.

\begin{figure}[t]
    \vspace{-4pt}
    \centering
    \includegraphics[width=0.95\columnwidth]{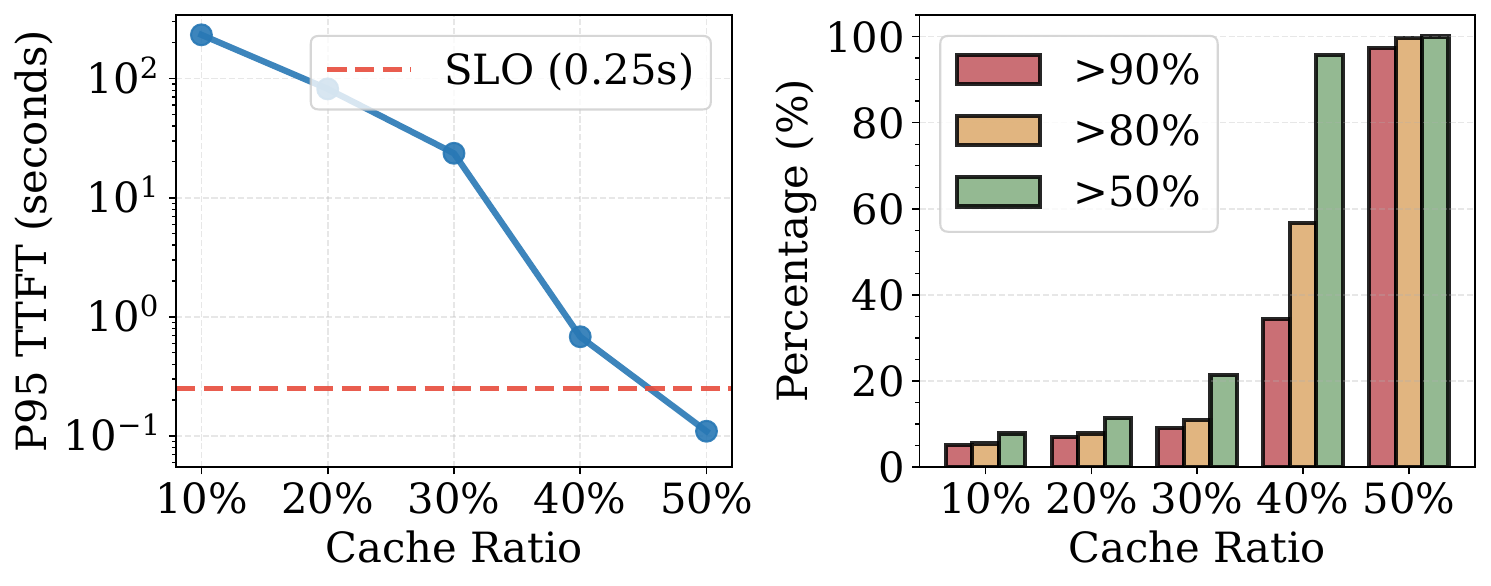}
        \vspace{-10pt}
    \caption{Impact of LoRA cache ratio on TTFT performance and SLO attainment. (\textbf{Left}) P95 TTFT under varying cache ratios, with SLO of 0.25 seconds. (\textbf{Right}) Percentage of LoRA adapters for which the fraction of requests meeting the TTFT SLO exceeds specific thresholds (50\%, 80\%, and 90\%).}
    \Description{Need to modify, just a simplified diagram}
    \label{fig:motivation_cache_ratio}
    \vspace{-1cm}
    \vspace{18pt}
\end{figure}

\noindent\textbf{Issue 1: a low cache capacity leads to excessive TTFT\footnote{In multi-LoRA serving with PD disaggregation, we focus on the latency of generating the first token in the decode phase, as discussed in Section~\ref{sec:background_llm_inference}. Accordingly, we redefine TTFT for a request as the sum of its queueing delay and the time required by the decoding engine to produce the first output token, explicitly excluding the prefill phase. This definition isolates the performance impact of decode-time LoRA serving and directly reflects user-perceived service quality.}, directly degrading service quality.} When the required LoRA adapter is not resident in GPU memory, incoming requests must wait until in-flight executions complete and GPU memory becomes available to load its adapter. This introduces additional queuing and loading delays before decoding can begin. Since TTFT is highly sensitive to such delays, even moderate cache misses can significantly inflate tail latency, resulting in poor SLO compliance. As shown in Figure~\ref{fig:motivation_cache_ratio}, the P95 TTFT reaches hundreds of seconds when the LoRA cache ratio (defined as the cache capacity divided by the total number of LoRAs) is low. As cache ratio increases, the P95 TTFT drops sharply. Further analysis of adapter-level compliance reveals a consistent trend: a larger cache ratio allows significantly more LoRA adapters to exceed high SLO satisfaction thresholds (e.g., >80\% or >90\% of their requests).

\begin{figure}[t]
    \vspace{-4pt}
    \centering
    \includegraphics[width=0.95\columnwidth, height=3.8cm]{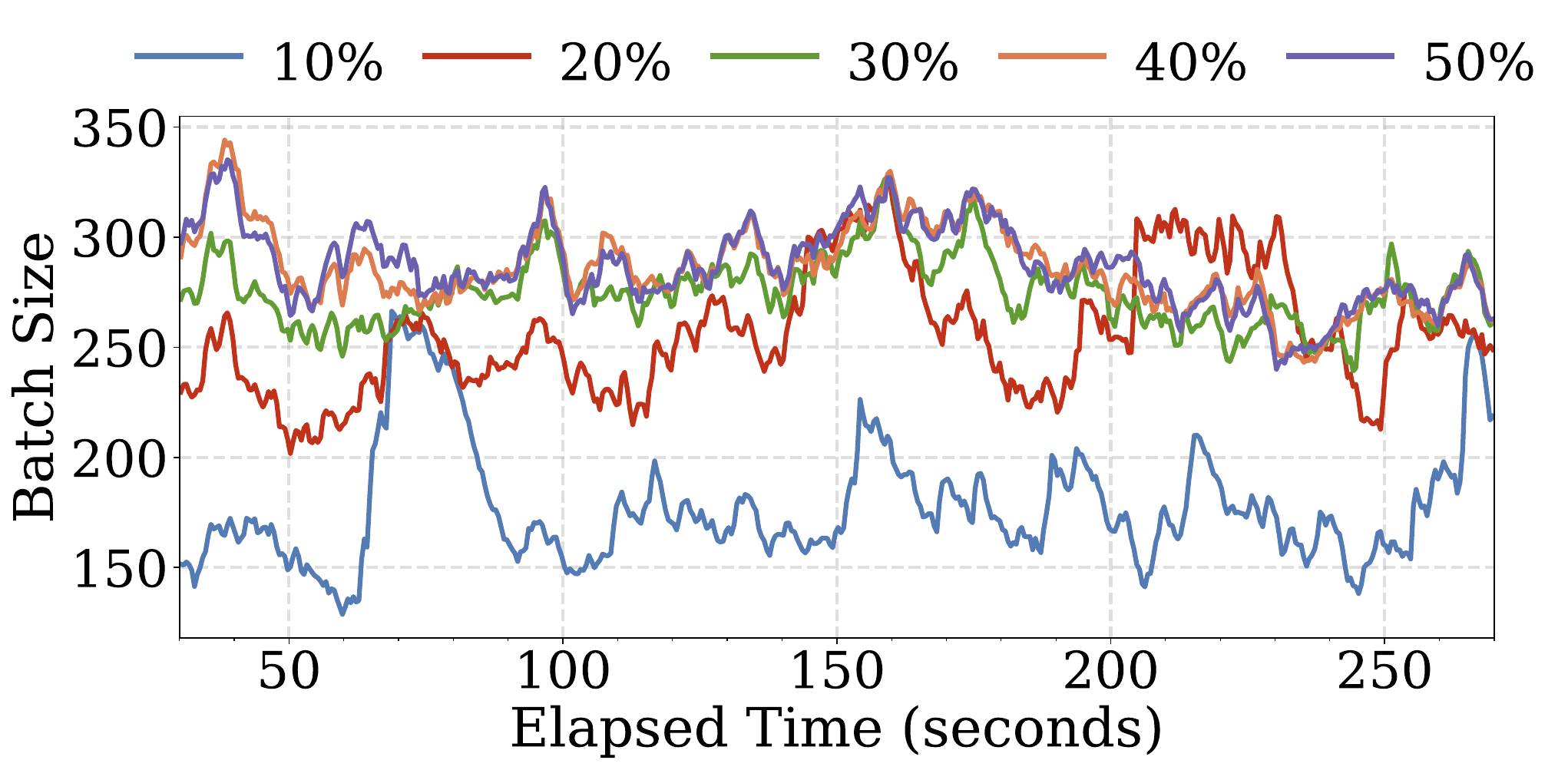}
        \vspace{-10pt}
    \caption{Impact of LoRA cache ratio on batch size. Measurements are collected during the steady-state interval (30–270s) of the 300s experiment.}
    \Description{Need to modify, just a simplified diagram}
    \label{fig:motivation_batch_size}
    \vspace{-12pt}
\end{figure}

\noindent\textbf{Issue 2: an insufficient cache capacity reduces the effective batch size, lowering hardware utilization.} 
Requests targeting uncached LoRA adapters cannot be admitted into the execution engine and must remain in the waiting queue. As a result, the engine operates with fewer requests than its configured batch size limit, even when sufficient compute resources are available. This underutilization is particularly detrimental during the decode phase, which requires large batch sizes to saturate memory bandwidth. Figure~\ref{fig:motivation_batch_size} illustrates the batch size observed over time. Request arrivals follow a Poisson process, causing the batch size to fluctuate dynamically. When cache ratio is low, the batch size remains consistently small because only a limited number of adapters can reside on GPUs at any time. Moreover, the batch size exhibits high variance because request admission is strictly constrained by the currently resident adapters.

\chen{These cache-capacity constraints are not unique to MoE models: dense models can also encounter them when their active adapter working sets exceed local cache capacity, although larger expert-specific adapters make MoE models particularly susceptible.} As discussed in Section~\ref{sec:introduction}, neither scaling out nor scaling up fundamentally resolves the limitations of LoRA serving under the \emph{coupled} architecture. While they differ in trade-offs (e.g., memory efficiency vs. communication overhead), both remain constrained by the same root cause: the tight coupling between LoRA adapters and base model execution. This coupling forces the scaling of lightweight, dynamic adapters to be entangled with the heavyweight base model, fundamentally limiting system scalability and efficiency.
\section{An Overview of InfiniLoRA}

To break the scalability limitations of the coupled architecture, we propose \textbf{InfiniLoRA}, a disaggregated LoRA serving system that decouples LoRA adapters from LLM instances. This design enables LoRA adapters to be shared across instances and allows LoRA cache capacity to scale independently of base-model execution.

\begin{figure}[tbp]
    \vspace{-4pt}
    \centering
    \begin{subfigure}{0.95\linewidth}
        \centering
        \includegraphics[width=\linewidth]{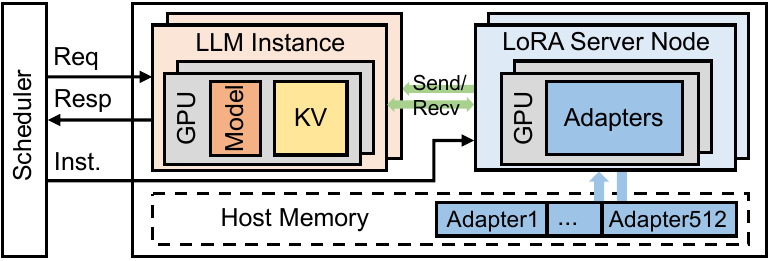}
        \vspace{-13pt}
        \caption{Disaggregated architecture.}
        \label{fig:system_overview}
    \end{subfigure}
    \begin{subfigure}{0.95\linewidth}
        \centering
        \includegraphics[width=\linewidth]{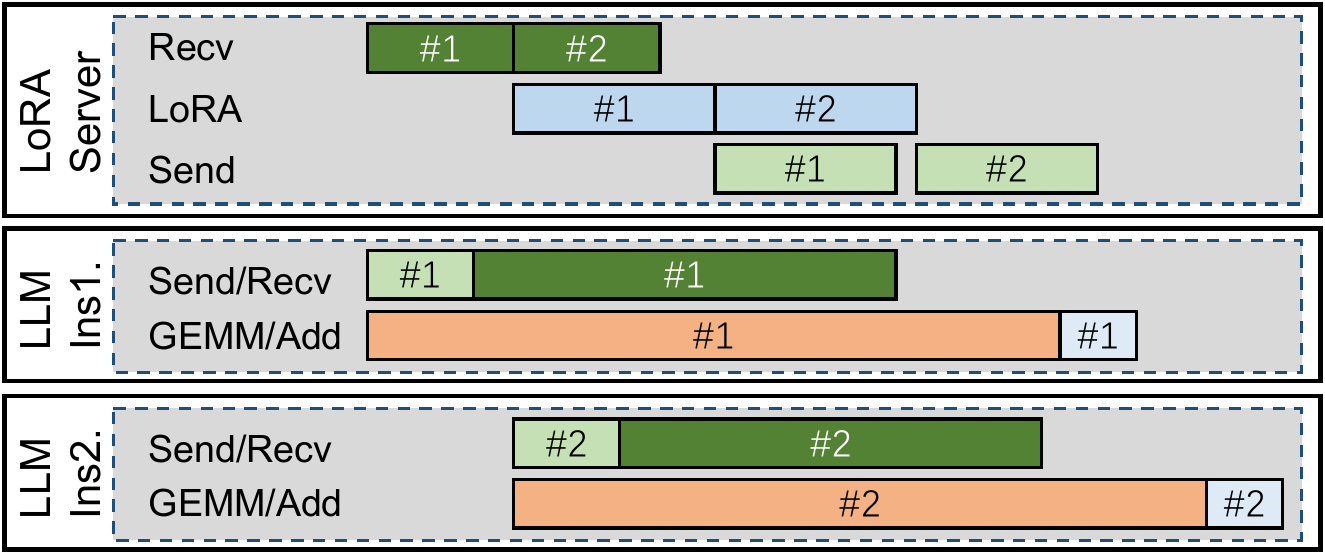}
        \vspace{-13pt}
        \caption{Execution workflow: LLM instances send activations to the LoRA Server and receive the computed results. This communication is overlapped with local GEMM computations, followed by a final addition of the two outputs. \chen{Gate/up synchronize before the intermediate activation; down forms a second stage with its own overlap and synchronization.}}
        \label{fig:execution_workflow}
    \end{subfigure}
        \vspace{-10pt}
    \caption{An overview of InfiniLoRA.}
    \Description{Need to modify, just a simplified diagram}
    \label{fig:overview}
    \vspace{-4pt}
\end{figure}

Figure~\ref{fig:system_overview} presents the architecture of InfiniLoRA. Unlike coupled designs, InfiniLoRA manages LoRA adapters in a dedicated, potentially multi-node LoRA Server, while LLM instances remain LoRA-free and execute the base model using existing optimizations. During inference, each LLM instance performs base-model computation and offloads LoRA-related computation by sending activations to the LoRA Server, which applies the corresponding LoRA transformations and returns updated activations as in Figure~\ref{fig:execution_workflow}. This interaction occurs at each MoE layer, forming a fine-grained remote execution pipeline.

\noindent\textbf{Challenges and Opportunities.}
Disaggregation fundamentally introduces a new bottleneck: communication between LLM instances and the LoRA Server. However, it also exposes a key opportunity, \emph{dual-level pipelining}, to hide this overhead. Specifically, from the LLM instance perspective, LoRA communication can be overlapped with MoE execution, while from the server perspective, communication from one instance can be overlapped with computation from others. This dual-level overlap transforms disaggregation from a communication-bound design into a pipeline-parallel execution model. In essence, InfiniLoRA converts remote LoRA execution from a latency overhead into an opportunity for cross instances pipeline parallelism.
In our design, we treat the upgate projection and the down projection of each expert FFN layer as two separate stages.
Figure~\ref{fig:execution_workflow} illustrates the execution workflow of one such stage.
Within the first upgate stage, the base GEMMs are overlapped with activation transfer, LoRA computation, and result transfer.
The base and LoRA outputs are synchronized and summed separately for gate and up. The resulting gate output passes through the nonlinear activation and is multiplied element-wise by the resulting up output to form the intermediate activation.
The subsequent down-projection stage uses this intermediate activation and repeats the overlap, synchronization, and addition.

Realizing this pipelined execution requires addressing three tightly coupled challenges. First, to enable effective overlap, the system must carefully design LoRA data placement and parallelism across GPUs, as it directly determines communication volume, computation parallelism, and synchronization scope under the M-to-N interaction between instances and the server. Second, to preserve the benefits of overlap, remote LoRA execution and communication must be optimized along the decode-time critical path to avoid degrading TTFT and TPOT. Third, to sustain pipeline efficiency under dynamic workloads, the LoRA Server must be provisioned in an SLO-aware manner to balance latency guarantees and resource utilization. InfiniLoRA addresses these challenges through parallelism-aware execution and SLO-driven resource provisioning (Section~\ref{sec:LoRAServerDesign}), as well as critical-path optimization (Section~\ref{sec:CriticalPathOptimization}).

\begin{figure*}[htbp]
    \vspace{-4pt}
    \centering
    \begin{subfigure}{0.45\textwidth}
        \centering
        \includegraphics[width=\linewidth]{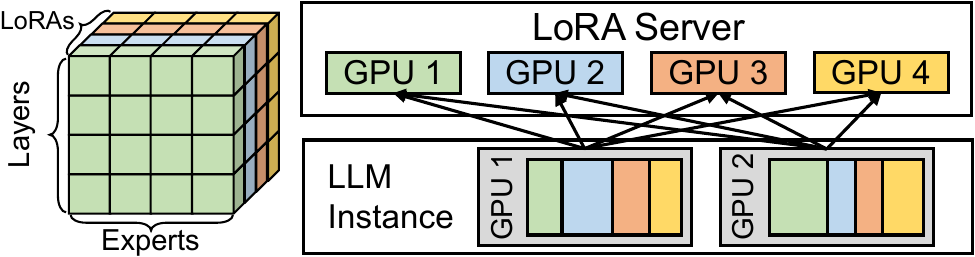}
        \vspace{-13pt}
        \caption{Data parallel.}
        \label{fig:data_layout_dp}
    \end{subfigure}
    \begin{subfigure}{0.45\textwidth}
        \centering
        \includegraphics[width=\linewidth]{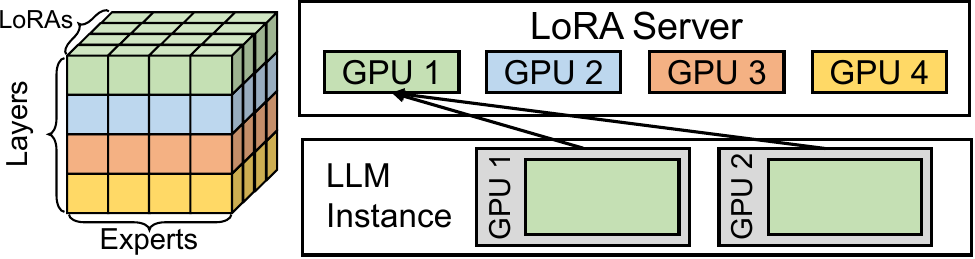}
        \vspace{-13pt}
        \caption{Pipeline parallel.}
        \label{fig:data_layout_pp}
    \end{subfigure}
    \begin{subfigure}{0.45\textwidth}
        \centering
        \includegraphics[width=\linewidth]{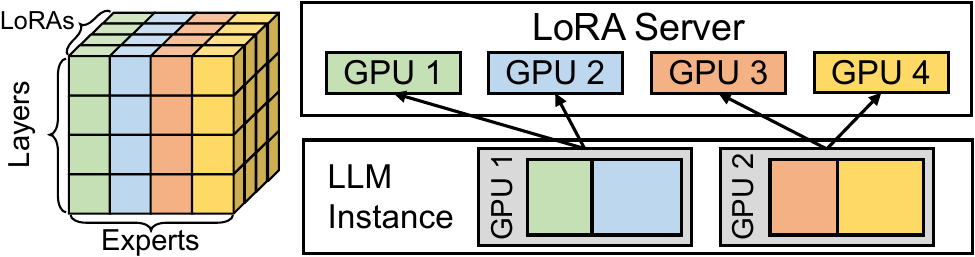}
        \vspace{-13pt}
        \caption{Expert parallel.}
        \label{fig:data_layout_ep}
    \end{subfigure}
    \begin{subfigure}{0.45\textwidth}
        \centering
        \includegraphics[width=\linewidth]{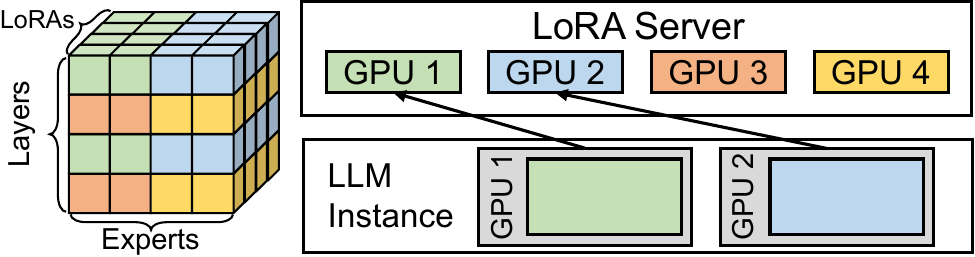}
        \vspace{-13pt}
        \caption{Hybrid parallel.}
        \label{fig:data_layout_hp}
    \end{subfigure}
        \vspace{-8pt}
    \caption{LoRA adapter placement strategies across server GPUs. The three-dimensional block represents the adapter space, with axes corresponding to LoRA adapters, layers, and experts. Each color indicates the server GPU (GPU~1--4) to which a partition of adapters is assigned. Arrows depict the activation data flow between client GPUs and server GPUs.}
    \label{fig:data_layout}
    \vspace{-4pt}
\end{figure*}

\noindent\textbf{Remarks.}
\chen{We target multi-LoRA serving workloads whose active adapter working sets exceed local adapter-cache capacity.
We focus on MoE models because expert-specific adapters cause adapter memory requirements to grow with the number of experts, amplifying cache-capacity pressure. 
The same capacity-pooling principle can also benefit dense models when their active adapter working sets exceed local cache capacity. 
The resulting performance benefits depend on the balance between reduced adapter-related admission delays and the additional overhead of remote execution.}

\chen{Our current implementation and evaluation focus on LoRA adapters for the gate, up, and down projections of MoE expert FFNs. Attention adapters do not scale with the number of experts, and their Q/K/V/O projections offer shorter overlap windows during decoding, making expert-FFN adapters the primary target of our disaggregation design. For mixed attention/FFN configurations, a potential extension is to retain attention adapters locally while disaggregating expert-FFN adapters. Evaluating this extension and disaggregated execution for attention configurations is left to future work.}

\chen{InfiniLoRA assumes a fixed deployment and does not support runtime elasticity or transparent fault recovery.
A GPU or node failure requires restarting or reconfiguring the affected deployment, and newly available nodes are incorporated only after reconfiguration. 
We adopt restart-based recovery because each LoRA Server operates within a bounded serving domain.
Communication and synchronization overhead constrain its scale (Sections 4), while its processing capacity limits the number of LLM instances it can serve within a layer-latency window. 
Recovery therefore operates within this bounded domain rather than across an arbitrarily large cluster. Recovery requires explicit handling of interrupted requests, repopulation of adapter caches, and re-establishment of affected client connections.}
\section{LoRA Server Design: Parallelism-Aware Execution and SLO-Driven Provisioning}
\label{sec:LoRAServerDesign}


\subsection{Parallelism-Aware LoRA Execution}
\label{sec:lora_execution}

Given an MoE model with $l$ layers, $e$ experts per layer, $n$ LoRA adapters, and a LoRA Server with $m$ GPUs, we organize LoRA weights as a three-dimensional tensor of size $n \times l \times e$, where each element corresponds to the LoRA weights of one expert in one layer for one adapter (Figure~\ref{fig:data_layout}). For brevity, we refer to the GPUs of an LLM instance and the LoRA Server as \emph{client} and \emph{server} GPUs, respectively.

Consider an LLM instance with $p$ GPUs and per-instance batch size $b$, using expert parallelism of degree $p$. Each request activates the top-$k$ experts per layer, so one decoding step generates $b \times k$ LoRA work units for each layer. Each work unit sends one activation vector to the LoRA Server and receives one updated activation vector in return. For clarity, we normalize both communication and computation volume by the size of one activation vector, so the total per-layer workload is $b \times k$.

We compare four parallelization strategies using the metrics in Table~\ref{tab:layout_comparison}: \emph{Comm Peer Volume} (average communication volume per active client--server GPU pair), \emph{Comm Peer Count} (number of client GPUs communicating with an active server GPU), \emph{Compute Volume} (LoRA computation assigned to an active server GPU), and \emph{Sync Scope} (number of server GPUs synchronized for one layer).

\noindent\textbf{LoRA Data Parallel (DP).}
A straightforward design is to shard adapters evenly across the $m$ server GPUs (Figure~\ref{fig:data_layout_dp}). Since requests in the same batch may target adapters placed on any server GPU, each client GPU may send activations to all server GPUs. Thus, each server GPU may communicate with all $p$ client GPUs, and the synchronization scope spans all $m$ server GPUs.

Each client GPU generates, on average, $\frac{b \times k}{p}$ work units per layer. With adapters uniformly distributed across $m$ server GPUs, these work units are further spread across all servers, yielding an average communication volume of $\frac{b \times k}{p \times m}$ per active client--server pair and an average compute volume of $\frac{b \times k}{m}$ per server GPU. In practice, skewed adapter popularity may still cause communication and compute imbalance despite uniform placement.

\noindent\textbf{LoRA Pipeline Parallel (PP).}
An alternative design is to partition LoRA weights by layer and place all adapters of one layer on the same server GPU, distributing the $l$ layers evenly across the $m$ server GPUs (Figure~\ref{fig:data_layout_pp}).\footnote{A single GPU typically has sufficient memory to hold all adapters for one layer. If not, multiple GPUs can be grouped for that layer. For simplicity, we assume one GPU per layer.}
For a given layer, all $p$ client GPUs send their activations to the single server GPU hosting that layer. Therefore, the communication peer count is $p$, the synchronization scope is $1$, the communication volume per client--server pair is $\frac{b \times k}{p}$, and the compute volume on the active server GPU is $b \times k$.

PP eliminates inter-server synchronization for a layer and avoids server-side load imbalance within that layer. It also allows different LLM instances to process different layers on different server GPUs. However, it concentrates all communication and computation of a layer on one server GPU, which can amplify critical-path latency.

\noindent\textbf{LoRA Expert Parallel (EP).}
In EP, we partition LoRA weights by expert and distribute the $e$ experts evenly across the $m$ server GPUs (Figure~\ref{fig:data_layout_ep}), so each server GPU hosts the adapters of $\frac{e}{m}$ experts. Since any of the top-$k$ experts may be activated, one layer may involve all $m$ server GPUs, giving a synchronization scope of $m$.

Assuming aligned expert partitioning, each client GPU owns $\frac{e}{p}$ experts. A server GPU only receives activations from client GPUs that own overlapping experts, so the communication peer count is $\max(\frac{p}{m}, 1)$. The $b \times k$ work units are evenly distributed across the $m$ server GPUs on average, yielding a compute volume of $\frac{b \times k}{m}$ per server GPU and a communication volume of $\frac{b \times k}{\max(p,m)}$ per active client--server pair.
Compared with PP, EP alleviates per-GPU compute concentration, but it enlarges the synchronization scope and suffers from fragmented communication and expert-level imbalance. 
\chen{Because the base-expert and LoRA computations follow the same token routing, a hot expert increases the workload on both paths. Although routing skew can lead to uneven LoRA workloads across server GPUs, the longer base-expert computation can also provide a larger window for hiding remote LoRA execution. Under the evaluated workloads, we did not observe expert-routing imbalance substantially impairing this overlap.}

\begin{table}[tbp]
\centering
\caption{Comparison of LoRA execution parallelization strategies.
$b$, $k$, $p$, $m$: per-instance batch size, expert routing top-$k$, GPU counts for an LLM instance and the LoRA Server. $x$, $y$: hybrid parallelism configuration.}
\label{tab:layout_comparison}
\vspace{-10pt}
\resizebox{\columnwidth}{!}{%
\begin{tabular}{ccccc}
\toprule
\makecell{\textbf{Parallel} \\ \textbf{Strategy}} & 
\makecell{\textbf{Comm} \\ \textbf{Peer Volume}} & 
\makecell{\textbf{Comm} \\ \textbf{Peer Count}} & 
\makecell{\textbf{Compute} \\ \textbf{Volume}} & 
\makecell{\textbf{Sync} \\ \textbf{Scope}} \\
\midrule
$DP$ & $\frac{b\times k}{p\times m}$ & $p$ & $\frac{b\times k}{m}$ & $m$ \\
$PP$ & $\frac{b\times k}{p}$ & $p$ & $b\times k$ & $1$ \\
$EP$ & $\frac{b\times k}{\max(p,m)}$ & $\max(\frac{p}{m},1)$ & $\frac{b\times k}{m}$ & $m$ \\
$EP_x\text{-}PP_y$ & $\frac{b\times k}{\text{max}(p,x)}$ & $\text{max}(\frac{p}{x},1)$ & $\frac{b\times k}{x}$ & $x$ \\
\bottomrule
\end{tabular}%
}
\vspace{-14pt}
\end{table}

\noindent\textbf{LoRA Hybrid Parallel.}
To balance synchronization overhead, communication fragmentation, and critical-path load, we combine expert and pipeline parallelism. We denote the configuration as $EP_x\text{-}PP_y$, where $x$ is the expert-parallel degree, $y$ is the number of pipeline stages, and $x \times y = m$. Figure~\ref{fig:data_layout_hp} shows the case of $x=2, y=2$. We interleave layers across stages rather than assigning contiguous layers, which helps reduce LoRA loading overhead (e.g., assigning Layers~1 and~3 to GPUs~1--2, and Layers~2 and~4 to GPUs~3--4).

For one layer, only the $x$ server GPUs in its expert group participate, so the synchronization scope is $x$. Each active server GPU communicates with $\max(\frac{p}{x}, 1)$ client GPUs, yielding a communication volume of $\frac{b \times k}{\max(p,x)}$ per active client--server pair and a compute volume of $\frac{b \times k}{x}$ per active server GPU. Increasing $x$ reduces per-GPU compute load but enlarges synchronization scope and may worsen communication fragmentation and expert imbalance; decreasing $x$ has the opposite effect. In practice, we tune $x$ and $y$ offline, and setting $x$ to the number of intra-node GPUs is often a good default (Section~\ref{sec:exp_scale_server}).

\noindent\textbf{Summary.}
The above analysis reveals that no single strategy dominates across all metrics. Data parallelism is straightforward but incurs a large synchronization scope, LoRA-level load imbalance, and fragmented communication and computation. Pipeline parallelism minimizes synchronization but concentrates all work on a single GPU, amplifying critical-path latency. Expert parallelism distributes computation more evenly but enlarges synchronization scope and is susceptible to fragmented communication and expert-level imbalance.

In contrast, hybrid parallelism enables a flexible trade-off among these factors. By tuning the expert-parallel degree $x$ and pipeline depth $y$, it balances synchronization scope, communication locality, and per-GPU compute load. This makes it well-suited for disaggregated LoRA serving, where minimizing critical-path latency while maintaining high resource utilization is essential.

\subsection{SLO-Driven LoRA Server Resource Provisioning}
\label{sec:resource_provision}

\noindent\textbf{Satisfying the Tail TTFT SLO.}
\textit{From TTFT to Immediate Admissibility.}
In low-latency inference systems, request queuing is the dominant cause of TTFT SLO violations. We formalize this through the \emph{Immediate Admissibility Rate}~(IAR): a request is \emph{immediately admissible} if its target LoRA adapter is already resident in GPU memory or can be instantly loaded into an available slot. 
\chen{Note that this definition does not account for queuing caused by saturated LLM instances or full batches.}
Under our simplified analysis, which abstracts away batch-capacity-induced queueing and treats immediate admissibility as the dominant factor of TTFT, satisfying the P95 TTFT SLO can be approximated by requiring IAR $\geq$ 95\%.
Given the invocation probability distribution $\{p_i\}_{i=1}^{N}$ over $N$ adapters, the global batch size $LB = L \cdot B$ ($L$: number of LLM instances, $B$: per-instance batch size), and a target IAR $\alpha \in (0,1]$, our goal is to find the minimum cache capacity:
\begin{equation}\label{eq:objective}
  M^{*} \;=\; \min\;\bigl\{\, M \in [N]
        \;\bigm|\; \text{IAR}(M) \geq \alpha \,\bigr\}
\end{equation}

\textit{Probabilistic Modeling.}
We model the system in steady state for a LoRA Server with cache capacity \(M\). For adapter \(i\), the expected access count in a global batch is \(\lambda_i = LB \cdot p_i\). To obtain a tractable approximation, we model this count as \(\text{Poisson}(\lambda_i)\) and treat adapter \(i\) as \emph{resident} if its count exceeds a threshold \(\tau^{*}\). The resulting residency probability is
\begin{equation}\label{eq:residency}
  q_i \;=\; \Pr\!\bigl[\text{Poisson}(\lambda_i) > \tau^{*}\bigr]
      \;=\; 1 - \sum_{k=0}^{\tau^{*}}
            \frac{\lambda_i^{k}\,\exp(-\lambda_i)}{k!}\, .
\end{equation}
We choose \(\tau^{*}\) numerically such that \(\sum_{i=1}^{N} q_i \approx M\). For tractability, we further approximate the residency events of different adapters as independent Bernoulli variables with marginals \(\{q_i\}_{i=1}^{N}\), while noting that a fixed-capacity cache induces correlations in the exact system.

For an incoming request targeting adapter~$i$, it avoids adapter-capacity-induced queuing in two cases: \emph{(i)}~a direct cache hit (probability $q_i$), or \emph{(ii)}~a cache miss but a free slot exists among the $M$ positions.
Let $P_{\text{free}}(i)$ denote the probability that the remaining $N-1$ adapters collectively occupy at most $M-1$ slots. Since
each adapter $j \neq i$ independently resides in the cache with probability $q_j$, the total number of occupied slots is a sum of $N-1$ independent Bernoulli random variables.
$P_{\text{free}}(i)$ is then the probability that this sum does not exceed $M-1$.
Aggregating over all adapters, the overall IAR is:
\begin{equation}\label{eq:iar}
  \text{IAR}(M)
  \;=\; \sum_{i=1}^{N} p_i \,\Bigl[\,
        q_i \;+\; (1-q_i)\;\cdot\;P_{\text{free}}(i)
        \,\Bigr]
\end{equation}

\textit{Solution Procedure.}
Algorithm~\ref{alg:ttft} solves Problem~\eqref{eq:objective} by incrementally testing $M = 1, 2, \ldots$ and returning the first value satisfying the IAR target. For each candidate~$M$, the algorithm proceeds in three stages.
First it solves $\sum q_i = M$ for $\tau^{*}$ via binary search (line~3), and obtains all residency probabilities $\{q_i\}$ (line~4).
Then it computes $P_{\text{free}}(i)$ for each adapter via dynamic programming and accumulates the per-adapter contribution to IAR (lines~6--16). Finally it returns $M^{*} = M$ once $\text{IAR} \geq \alpha$ (line~17).
Naturally, the minimum GPU memory budget for the LoRA Server is $M^{*} \times \text{Mem}_{\text{LoRA}}$, where $\text{Mem}_{\text{LoRA}}$ is the per-adapter footprint.

\begin{algorithm}[t]
\caption{Minimum Cache Size for a Target IAR}
\label{alg:ttft}
\begin{algorithmic}[1]
\REQUIRE LoRA invocation probabilities $\{p_i\}_{i=1}^{N}$, global batch size $LB$, target IAR $\alpha$
\ENSURE  Minimum cache size $M^{*}$
\STATE Compute $\lambda_i \gets LB \cdot p_i$ for all $i \in [N]$
\FOR{$M = 1$ \textbf{to} $N$}
    \STATE $\tau^{*} \gets \text{\textsc{BinarySearch}}(\tau \text{ s.t.\ } \sum_{i=1}^{N} Q(\lambda_i,\tau) = M)$
    \COMMENT{$Q(\lambda,\tau)=\Pr[\text{Poisson}(\lambda)>\tau]$}
    \STATE $q_i \gets Q(\lambda_i,\,\tau^{*})$ for all $i$
    \STATE $\text{IAR} \gets 0$
    \FOR{$i = 1$ \textbf{to} $N$}
        \STATE $\text{dp}[0] \gets 1$; \; $\text{dp}[1..N] \gets 0$
        \COMMENT{DP over Poisson-Binomial}
        \FOR{each $j \in [N] \setminus \{i\}$}
            \FOR{$k = N-1$ \textbf{down to} $1$}
                \STATE $\text{dp}[k] \gets \text{dp}[k]\cdot(1-q_j) + \text{dp}[k\!-\!1]\cdot q_j$
            \ENDFOR
            \STATE $\text{dp}[0] \gets \text{dp}[0]\cdot(1-q_j)$
        \ENDFOR
        \STATE $P_{\text{free}}(i) \gets \sum_{k=0}^{M-1}\text{dp}[k]$
        \STATE $\text{IAR} \gets \text{IAR} + p_i \cdot \bigl[q_i + (1-q_i)\cdot P_{\text{free}}(i)\bigr]$
    \ENDFOR
    \IF{$\text{IAR} \geq \alpha$}
        \RETURN $M^{*} \gets M$
    \ENDIF
\ENDFOR
\end{algorithmic}
\end{algorithm}

\noindent\textbf{Satisfying the Average TPOT SLO.}
We next address the TPOT SLO by deriving the computation
and communication resources required by the LoRA Server from two perspectives. 
From the perspective of \emph{LLM instance}, the global average TPOT SLO imposes a latency constraint:
\begin{equation}\label{eq:ffn_slo}
  T_{\text{recv}} + T_{\text{comp}} + T_{\text{send}}
  \;\leq\; \text{SLO}_{\text{FFN}}
\end{equation}
where $T_{\text{recv}}$, $T_{\text{comp}}$, $T_{\text{send}}$ denote the latencies incurred by the LoRA Server to receive activations, execute LoRA computation, and send back the results for a single LLM instance, respectively.
All three terms can be profiled and modeled as functions of the per-instance batch size $B$.
Additionally, $\text{SLO}_{\text{FFN}}$ represents the latency budget for the experts' Grouped GEMM, which is derived from the global average TPOT SLO.
From the \emph{LoRA Server} perspective, the LoRA Server must serve all $L$ LLM instances within the time window of a single layer, yielding the throughput constraint:
\begin{equation}\label{eq:layer_slo}
  \max(T_{\text{recv}},\;T_{\text{comp}},\;T_{\text{send}})
  \;\cdot\; L
  \;\leq\; \text{SLO}_{\text{Layer}}
\end{equation}
where $\text{SLO}_{\text{Layer}}$ denotes the total latency budget for one base model layer, encompassing both the attention and expert modules.
\chen{In summary, for $L$ LLM instances, let $S_{\min}(L)$ be the minimum server GPU count satisfying the cache requirement from Eq.~\eqref{eq:objective} for the tail TTFT SLO, determined by $L$, $B$, and adapter popularity, and the processing constraints in Eqs.~\eqref{eq:ffn_slo} and \eqref{eq:layer_slo}. 
With $g$ GPUs per instance and a total GPU budget $G$, feasible allocations satisfy $Lg + S_{\min}(L) \leq G$, rather than a fixed LLM-to-server ratio.}

\section{Critical-Path Optimization}
\label{sec:CriticalPathOptimization}

Section~\ref{sec:LoRAServerDesign} defines the execution structure of disaggregated LoRA serving, including its communication, computation, and synchronization pattern. This section then translates that structure into an efficient implementation by optimizing the critical path.

\noindent\textbf{Overlap Analysis} For a coarse-grained, bandwidth-based analysis, the MoE GEMM latency can be approximated as $T_{\mathrm{gemm}} = \frac{h \times d \times e}{BW_{\mathrm{mem}}}$,
where $h \times d$ is the size of one expert weight matrix and $e$ is the number of experts. In contrast, disaggregated LoRA execution only transfers activations to and from the LoRA Server, whose communication latency can be approximated as
$T_{\mathrm{comm}} = \frac{b \times k \times (h + d)}{BW_{\mathrm{comm}}}$,
where $b$ is the batch size and $k$ is the top-$k$ routing factor.
$BW_{\mathrm{mem}}$ and $BW_{\mathrm{comm}}$ here represent the GPU memory access bandwidth and the inter-node communication bandwidth, respectively.

Under typical serving configurations, $b \times k \times (h+d)$ is much smaller than $ h \times d \times e$, so the communication volume is substantially smaller than the MoE weight-access volume. As a result, although $BW_{\mathrm{mem}} \gg BW_{\mathrm{comm}}$, the communication latency can in principle be hidden behind MoE computation. For example, with $h{=}4096$, $d{=}16384$, $e{=}8$, $k{=}2$, $b{=}128$, $BW_{\mathrm{mem}}{=}4\,\mathrm{TB/s}$, and $BW_{\mathrm{comm}}{=}50\,\mathrm{GB/s}$, we still obtain $T_{\mathrm{comm}} < T_{\mathrm{gemm}}$. However, this theoretical overlap does not eliminate the need for careful optimization. 
\chen{This bandwidth estimate omits fixed transfer latency. Lower bandwidth or higher latency increases $T_{\mathrm{recv}}$ and $T_{\mathrm{send}}$ in Eqs.~\eqref{eq:ffn_slo}--\eqref{eq:layer_slo}, potentially violating the latency constraints and exposing communication.}
In practice, communication, LoRA computation, and weight loading all contribute to the decode-time critical path, and must therefore be carefully optimized to achieve high end-to-end performance.

\begin{figure}[tbp]
    \vspace{-4pt}
    \centering
    \begin{subfigure}{0.95\linewidth}
        \centering
        \includegraphics[width=\linewidth]{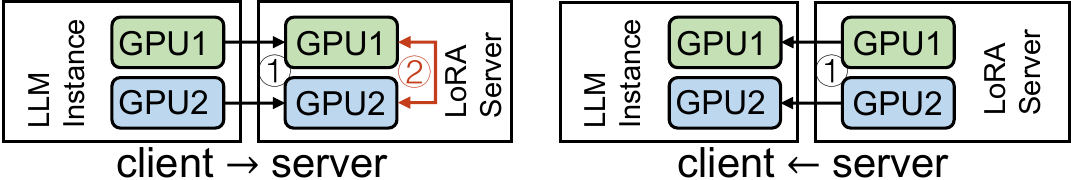}
        \vspace{-13pt}
        \caption{Push-based communication.}
        \label{fig:network_push_based}
    \end{subfigure}
    \begin{subfigure}{0.95\linewidth}
        \centering
        \includegraphics[width=\linewidth]{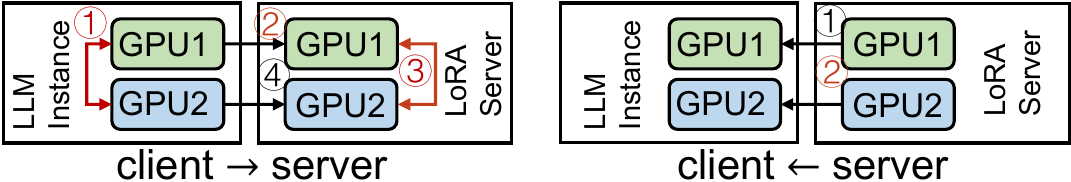}
        \vspace{-13pt}
        \caption{Pull-based communication.}
        \label{fig:network_pull_based}
    \end{subfigure}
        \vspace{-10pt}
    \caption{Design for client–server communication. Control signals are labeled red and data transfers are labeled black.}
    \Description{Need to modify, just a simplified diagram}
    \label{fig:network}
    \vspace{-10pt}
\end{figure}

\noindent\textbf{Host-Bypass Client-Server GPU Communication}
Although the above analysis confirms that LoRA communication can be overlapped with MoE computation, a low-latency, asynchronous communication mechanism between the LoRA server and LLM instances remains essential to fully realize this overlap in practice.
Unlike traditional collective communication, client–server interaction follows a dynamic, asynchronous pattern, where LoRA requests arrive independently from different instances.
This precludes NCCL P2P primitives, which rely on sender–receiver rendezvous. We instead adopt \texttt{IBGDA} (InfiniBand GPUDirect Async)~\cite{markthub2022improving}, enabling one-sided RDMA directly from GPU kernels. We support both \emph{push-based} and \emph{pull-based} modes, and describe the client-to-server path as a representative example in Figure~\ref{fig:network}.

\emph{Push-based Mode.}
Client GPUs directly write activations into preallocated buffers on target server GPUs via one-sided RDMA writes (\ding{172}). A leader server GPU runs a persistent kernel that polls these buffers and, upon detecting a completed write, broadcasts the client ID to other server GPUs to coordinate processing (\ding{173}). This avoids sender–receiver synchronization and enables low-latency batch admission.

\emph{Pull-based Mode.}
In contrast, pull-based communication requires clients to first coordinate locally (\ding{172}) and notify the server (\ding{173}), after which server GPUs must synchronize among themselves (\ding{174}) before issuing remote reads (\ding{175}). This introduces additional synchronization steps and network round trips. Our measurements show that for typical payloads (e.g., 4\,MB), pull-based communication incurs $2.63\times$ higher latency than its push-based counterpart.

A similar asymmetry applies to the server-to-client path: pull-based designs require clients to repeatedly poll remote completion states, incurring additional round-trip latency. Accordingly, InfiniLoRA adopts push-based communication in both directions to minimize decode-time latency.

\noindent\textbf{Hardware-Specialized LoRA Kernels}
LoRA computation differs from dense GEMM due to scattered memory access and fine-grained GEMM/GEMV operations, leading to low arithmetic intensity and poor hardware utilization. As communication already occupies part of the decode-time critical path, unoptimized LoRA computation can easily become a bottleneck.
Building on SGMV~\cite{chen2023punicamultitenantloraserving} and BGMV~\cite{sheng2023slora}, we design hardware-specialized kernels tailored to modern GPUs. 

We leverage GPU-specific features including wgmma, TMA (Tensor Memory Accelerator), warp specialization, and dynamic register reassignment.
For BGMV, where computational intensity is low, we adopt thread collaborative execution for computation instead of the heavier wgmma pipeline.
For SGMV, we apply the swapping-AB transformation (computing $A^{T}x^{T}$) to align tensor shapes with hardware constraints, enabling efficient use of \texttt{wgmma.m64n8k16}.
Additional optimizations include task scheduling and persistent kernels. We omit further details due to space constraints and will release the implementation.

\begin{figure}[tbp]
    \vspace{-4pt}
    \centering
    \includegraphics[width=0.95\columnwidth]{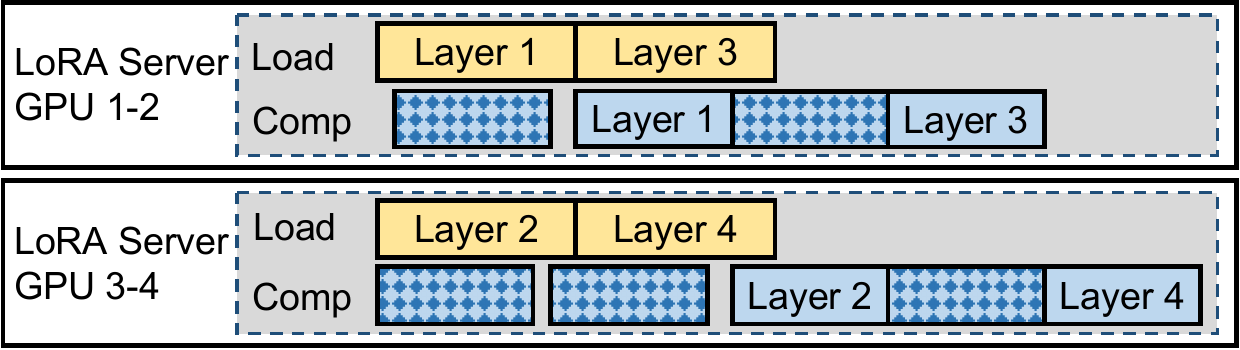}
        \vspace{-10pt}
    \caption{Layer-wise LoRA loading. Shaded blue blocks represent LoRA execution from any other LLM instance.}
    \Description{Need to modify, just a simplified diagram}
    \label{fig:lora_loading}
    \vspace{-6pt}
\end{figure}

\noindent\textbf{Layer-wise LoRA Loading} 
Dedicated server GPUs decouple adapter cache capacity from the memory available on LLM instances, allowing cache expansion without additional base-model copies.
This capacity is still bounded by the GPU memory of the LoRA Server, so cache misses cannot be eliminated and require on-demand loading of LoRA weights. To keep such loading from violating the TTFT SLO, we design a layer-wise loading strategy that works together with our hybrid parallel execution (Figure~\ref{fig:lora_loading}).

Layer-wise prefetching schedules future-layer loads ahead of execution. Our hybrid $EP_x\text{-}PP_y$ placement organizes server GPUs into $y$ pipeline groups, each containing $x$ GPUs.
Each layer's expert weights are sharded across the GPUs of its assigned group, allowing concurrent loading through multiple host-to-GPU paths and reducing the transfer volume handled by each GPU.
With multiple pipeline groups, successive layers are interleaved across groups, allowing different groups to load different layers of the same adapter concurrently. 
Together with expert sharding within each group, this enables parallel loading both within a layer and across layers. 
\nchen{For example, under the $EP_2\text{-}PP_2$ placement shown in Figure~\ref{fig:lora_loading}, one GPU group is assigned Layers~1 and~3, and the other Layers~2 and~4. Once Layer~1's shards are ready, the first group can execute it while prefetching Layer~3; the second group can concurrently load Layers~2 and~4.}

\chen{Out-of-band signaling further hides the loading latency: the scheduler instructs the LoRA Server to prefetch adapter weights before the LLM instance issues its first LoRA computation for Layer~1, which often eliminates cold-start stalls entirely.} 
With these optimizations, adapter loading does not violate the TTFT SLO in our setup with PCIe 5.0 (nearly 50 GB/s).

\section{Evaluation}

\aptLtoX{\begin{table}[tbp]
\centering
\caption{Model and LoRA configurations. Instance \#GPU denotes GPUs per LLM instance.}
\vspace{-8pt}
\resizebox{\columnwidth}{!}{%
\begin{tabular}{lccccc}
\hline
\textbf{Model} & 
\textbf{\#Layers} & 
\textbf{\#Experts} & 
\textbf{Top-k} & 
\begin{tabular}{@{}c@{}}\textbf{LoRA} \\ \textbf{Rank}\end{tabular} & 
\begin{tabular}{@{}c@{}}\textbf{Instance} \\ \textbf{\#GPU}\end{tabular} \\
\hline
GPT-OSS-20B~\cite{openai2025gptoss120bgptoss20bmodel} & 32 & 32 & 4 & 64 & 1 \\
Qwen3-30B-A3B~\cite{qwen3technicalreport}  & 48 & 128 &  8 & 32 & 2 \\
Mixtral-8x7B~\cite{jiang2024mixtralexperts}  & 32 & 8 & 2 & 64 & 2 \\
Scaled-MoE~\cite{zhu2025megascleinfer}  &  18 & 32 & 4 & 64 & 4 \\
DBRX~\cite{databricks2024dbrx} &  40 & 16 & 4 & 64 & 4 \\
\hline
\end{tabular}%
}
\label{tab:model_config}
\vspace{-5pt}
\end{table}}{\begin{table}[tbp]
\centering
\caption{Model and LoRA configurations. Instance \#GPU denotes GPUs per LLM instance.}
\vspace{-8pt}
\resizebox{\columnwidth}{!}{%
\begin{tabular}{lcccccc}
\toprule
\textbf{Model} & 
\textbf{\#Layers} & 
\textbf{\#Experts} & 
\textbf{Top-k} & 
\parbox{1.cm}{\centering \textbf{LoRA} \\ \textbf{Rank}} & 
\parbox{1.2cm}{\centering \textbf{Instance} \\ \textbf{\#GPU}} \\
\midrule
GPT-OSS-20B~\cite{openai2025gptoss120bgptoss20bmodel} & 32 & 32 & 4 & 64 & 1 \\
Qwen3-30B-A3B~\cite{qwen3technicalreport}  & 48 & 128 &  8 & 32 & 2 \\
Mixtral-8x7B~\cite{jiang2024mixtralexperts}  & 32 & 8 & 2 & 64 & 2 \\
Scaled-MoE~\cite{zhu2025megascleinfer}  &  18 & 32 & 4 & 64 & 4 \\
DBRX~\cite{databricks2024dbrx} &  40 & 16 & 4 & 64 & 4 \\
\bottomrule
\end{tabular}%
}
\label{tab:model_config}
\vspace{-0.3cm}
\vspace{-5pt}
\end{table}
}

\subsection{Evaluation Setup}

Our experiments are conducted on a four-node cluster. Each node is equipped with four NVIDIA Hopper GPUs (96~GB), 96 CPU cores, 2~TB of host memory, and four 400~Gb/s InfiniBand NICs, each directly attached to one GPU. GPUs within a node are connected via 900~GB/s NVLink. \chen{We evaluate InfiniLoRA on five MoE models summarized in Table~\ref{tab:model_config}, all with randomly initialized weights.} Unless otherwise specified, GPT-OSS-20B is evaluated with 8 GPUs, while all other models are evaluated with 16 GPUs. For Qwen3-30B-A3B, we use a reduced LoRA rank of 32 due to its fine-grained expert structure. 

\noindent\textbf{Workloads.}
Following prior work~\cite{chen2023punicamultitenantloraserving, sheng2023slora, li2025toppings, Iliakopoulou2025Chameleon}, we simulate a multi-tenant workload in which LoRA adapter popularity follows a Zipf distribution ($s=1.2$), calibrated to production observations in~\cite{zhu2025Cannikin}. 
\chen{Unless otherwise specified, each model experiment uses a global pool of 512 adapters, with request adapter IDs drawn according to this distribution. Arrivals follow a Poisson process, and we vary the request rate. Input/output length pairs are sampled from the same BurstGPT record~\cite{BurstGPT}, without filtering, truncation, or resampling.}

\noindent\textbf{Metrics.}
We report P95 TTFT to capture tail latency, especially under cache misses and queueing, and average TPOT to measure steady-state decoding efficiency. TTFT and TPOT SLOs are set to 0.25~s and 0.1~s, respectively, following prior work~\cite{zhu2025megascleinfer}. We further report the \emph{SLO Attainment Rate}~\cite{zhu2025Cannikin}: treating each adapter as an independent service, an adapter is considered compliant if more than 90\% of its requests satisfy both SLOs, and the attainment rate is the fraction of compliant adapters.

\noindent\textbf{Methods Under Study.}
InfiniLoRA is implemented as a distributed system consisting of a scheduler, multiple LLM instances built on vLLM~\cite{kwon2023vllm}, and a dedicated LoRA server. LLM instances communicate with the LoRA server through a lightweight connector, and all LoRA kernels are implemented in CUDA. By default, the LoRA server and LLM instances are deployed on separate nodes and communicate over InfiniBand.

We compare against S-LoRA~\cite{sheng2023slora} using its integrated vLLM implementation, which already supports GPU adapter caching. For each model, InfiniLoRA and S-LoRA are allocated the same total number of GPUs and use identical LLM-instance configurations; InfiniLoRA dedicates part of the budget to the LoRA server and uses the remainder for LLM instances, whereas S-LoRA uses all GPUs in the coupled serving architecture. Unless otherwise specified, S-LoRA allocates 50\% of the remaining GPU memory after loading model weights to LoRA cache and the other 50\% to KV cache. To study this trade-off, we also include a variant that allocates 40\% and 60\% of non-model GPU memory to the LoRA cache and KV cache, respectively, denoted \textit{S-LoRA w/ Less LoRA}. Both systems use the same scheduler described in Section~\ref{sec:lora_serving_background}. 
\chen{To assess length-based request ordering, we additionally evaluate an oracle variant, \textit{S-LoRA w/ SJF}, which uses Shortest-Job-First scheduling with oracle output lengths.}
For both InfiniLoRA and all S-LoRA variants, each LLM instance is pre-assigned a disjoint subset of adapters using a greedy load-balancing algorithm.

\chen{We discuss three closely related systems here. Toppings~\cite{li2025toppings} mitigates adapter-loading stalls by running LoRA computation on the CPU during prefill and switching to the GPU afterward. Since we target GPU decoding under strict latency SLOs, we do not compare against it directly. 
Chameleon~\cite{Iliakopoulou2025Chameleon} and Cannikin~\cite{zhu2025Cannikin} build on request scheduling and add further optimizations for multi-LoRA serving, such as adapter-aware cache management. Our baselines cover two of these building blocks: the vLLM-based S-LoRA baseline already provides adapter caching, and \textit{S-LoRA w/ SJF} uses oracle output lengths to examine length-based request ordering under limited LoRA cache capacity. Neither baseline represents the full cache-management and scheduling mechanisms of these systems.}
\chen{These systems optimize adapter caching within LLM-instance GPU resources, whereas InfiniLoRA expands adapter capacity using dedicated server GPUs without additional base-model copies. Their performance benefits can overlap while addressing different aspects of adapter caching.}

\subsection{Overall Performance Comparison}

\begin{figure*}[tbp]
    \vspace{-4pt}
    \centering
    \includegraphics[width=0.98\textwidth]{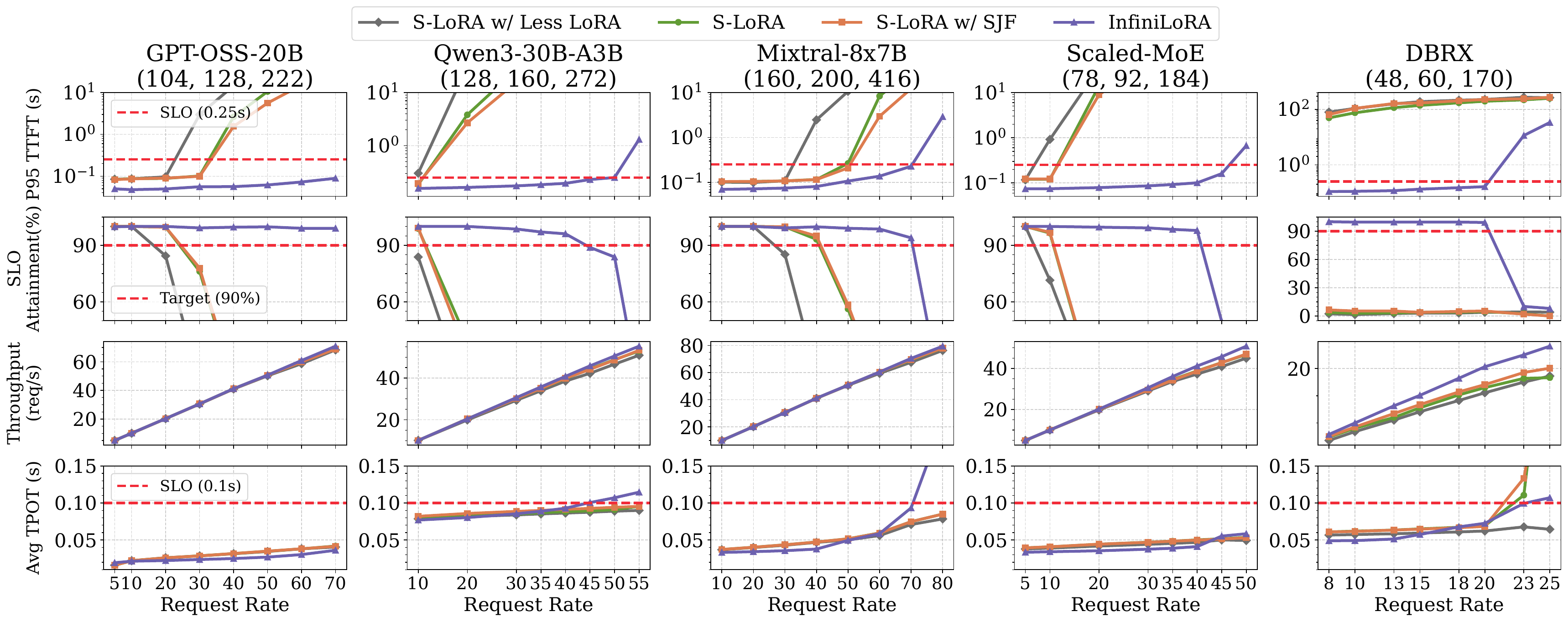}
        \vspace{-10pt}
    \caption{P95 TTFT, SLO attainment rate, throughput and average TPOT from top to bottom under varying loads. We compare InfiniLoRA against three baselines. The three values listed under each model name correspond to the LoRA cache capacity provided by S-LoRA w/ Less LoRA, S-LoRA (including w/ SJF) and InfiniLoRA, respectively.}
    \Description{Need to modify, just a simplified diagram}
    \label{fig:all_performance}
    \vspace{-0.3cm}
\end{figure*}

Figure~\ref{fig:all_performance} compares performance across models and request rates. For S-LoRA, given the LLM instance configurations in Table \ref{tab:model_config}, the number of instances is determined by dividing the total number of GPUs by the GPUs required per instance. InfiniLoRA follows an SLO-driven provisioning strategy to first determine number of GPUs for LoRA Server and then assigning the remaining GPUs to LLM instances. 
In most experiments, InfiniLoRA uses 8 GPUs across two nodes for the LoRA Server, which provides a good balance between communication, computation, and synchronization. Due to testbed limitations, the number of LLM instances cannot be further increased even though the LoRA Server has spare capacity; we therefore evaluate scalability separately in Section~\ref{sec:scale_num_instances}.

InfiniLoRA achieves an average $3.05\times$ increase in serviceable request rate across five models over S-LoRA while meeting both P95 TTFT and average TPOT SLOs. Moreover, InfiniLoRA improves the SLO attainment rate by an average of 54.0\% and 53.1\% compared to S-LoRA and S-LoRA w/ SJF, respectively. 
\chen{As the request rate increases, the performance of the baselines degrades rapidly. This is because higher request rates increase concurrency and can enlarge the set of adapters that must be resident at the same time, i.e., the adapter working set. 
When resident adapters cannot be reclaimed, requests that need additional adapters have to wait in the queue, which in turn increases tail TTFT.}
Against S-LoRA w/ Less LoRA, which allocates a smaller LoRA cache, the gains are even more pronounced: InfiniLoRA achieves a $4.56\times$ higher serviceable request rate and improves SLO attainment by 60.6\%.
\chen{Surprisingly, oracle SJF variant \textit{S-LoRA w/ SJF} offers limited gains under cache constraints.}
InfiniLoRA also improves throughput by 7.3\% on average and up to 24.7\% on DBRX compared to S-LoRA. Notably, InfiniLoRA attains higher throughput despite using fewer LLM instances, as the larger effective batch size enabled by increased LoRA cache capacity improves GPU utilization. 
\chen{At high request rates on Mixtral, InfiniLoRA exhibits higher TPOT than S-LoRA. This occurs because both systems cache a large number of adapters, but under our fixed testbed resources InfiniLoRA runs with only half the number of LLM instances. With comparable throughput, each InfiniLoRA instance therefore handles nearly twice the request load, increasing batch sizes and TPOT.}

\subsection{Scalability Evaluation}

\subsubsection{Scale with the Number of LLM Instances}
\label{sec:scale_num_instances}

\begin{figure}[tbp]
    \centering
    \begin{subfigure}{0.95\linewidth}
        \centering
        \includegraphics[width=\linewidth, height=2.1cm]{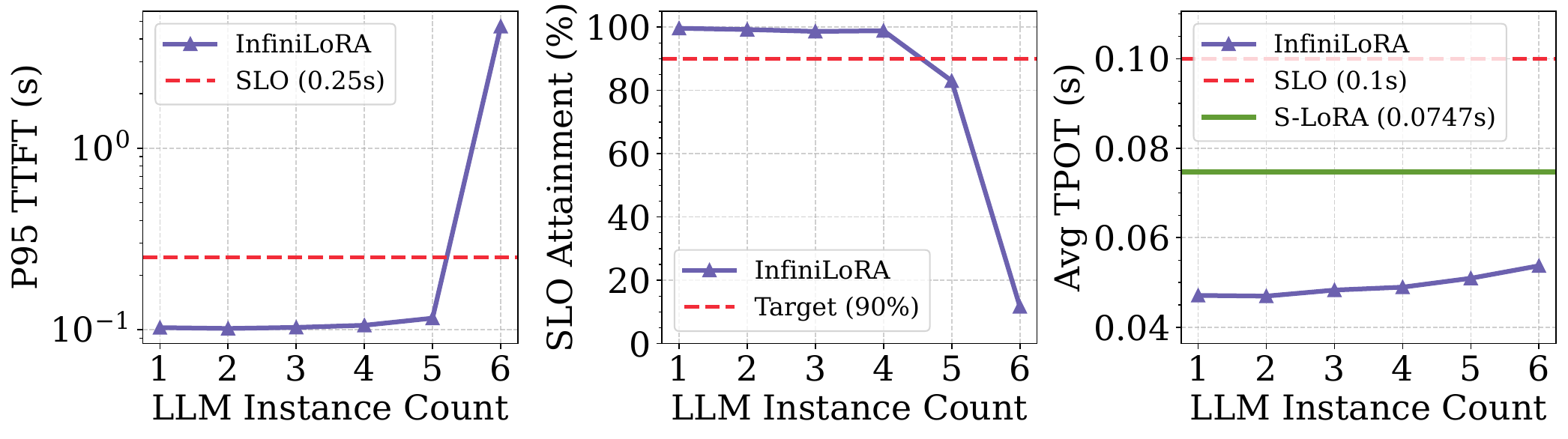}
        \vspace{-13pt}
        \caption{P95 TTFT, SLO attainment rate and average TPOT from left to right while scaling \#LLM instances.}
        \label{fig:scale_num_client}
    \end{subfigure}
    \begin{subfigure}{0.95\linewidth}
        \centering
        \includegraphics[width=\linewidth, height=3.4cm]{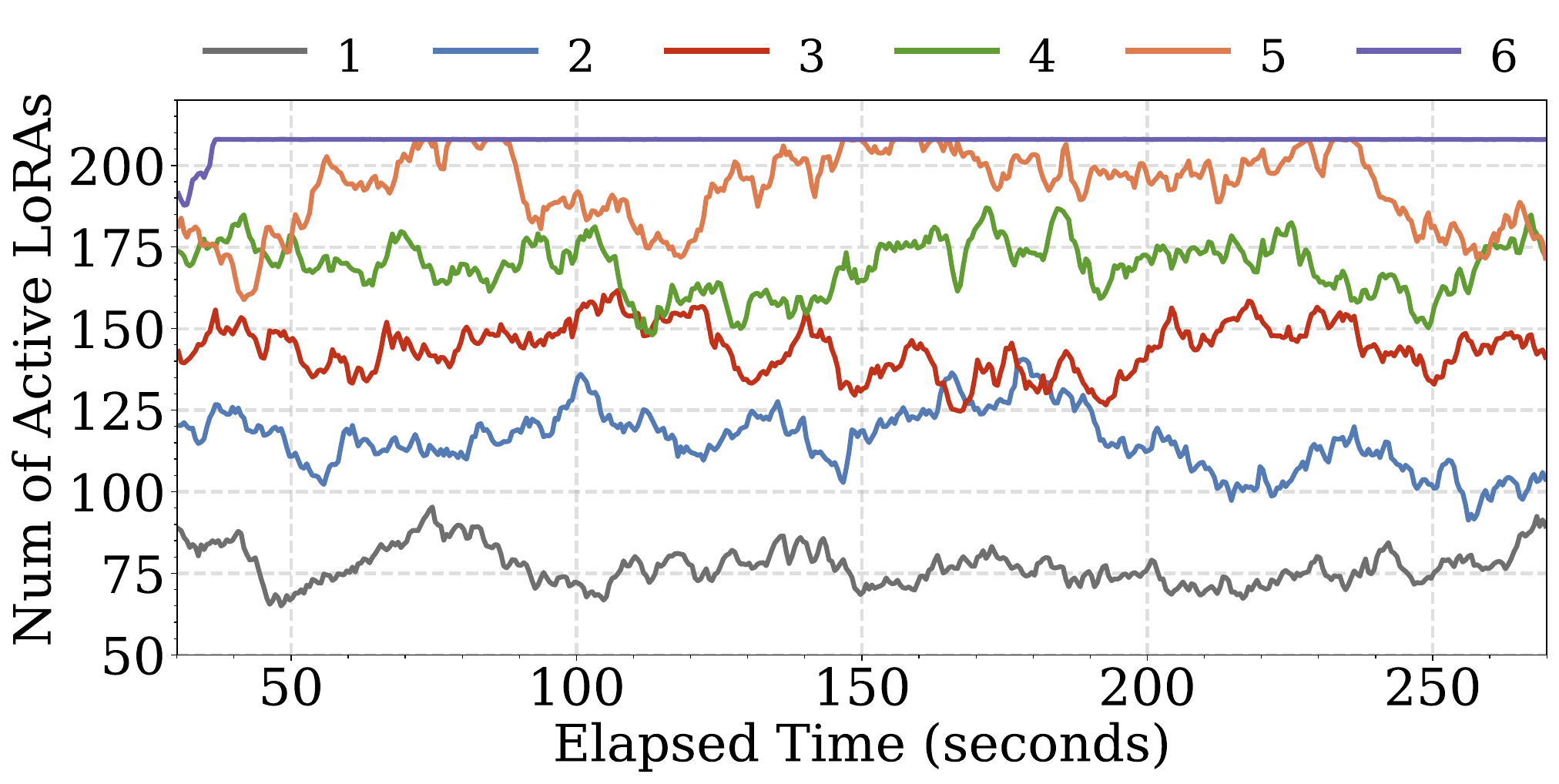}
        \vspace{-13pt}
        \caption{The evolution of active LoRA counts over time.}
        \label{fig:scale_num_client_lora_count}
    \end{subfigure}
        \vspace{-10pt}
    \caption{Performance of scaling \#LLM instances, configured with a request rate of 12 req/s per instance (keeping LoRA Server unchanged and using Mixtral-8x7B model).}
    \label{fig:scale_num_client_2fig}
    \vspace{-10pt}
\end{figure}

As shown in Figure~\ref{fig:scale_num_client_2fig}, we scale the number of LLM instances from 1 to 6, allocating two GPUs per instance to serve Mixtral model, while fixing the LoRA Server at 4 GPUs. We proportionally increase the aggregate request rate from 12 to 72 req/s to maintain constant per-instance load. Average TPOT remains relatively stable, increasing by only 10.5\% as the load on the LoRA Server gradually intensifies and still under SLO (0.1 s). Compared to S-LoRA (6 instances at 72 req/s), InfiniLoRA achieves lower TPOT because LoRA computation is largely overlapped with base-model execution, whereas S-LoRA executes LoRA serially. 
A 4-GPU LoRA Server thus provides sufficient compute throughput for six LLM instances.
However, both P95 TTFT and SLO attainment degrade sharply when the number of instances reaches six. 
Increasing aggregate concurrency expands the required adapter working set; once active adapters saturate the fixed server cache, requests needing additional adapters must wait for space to be reclaimed (Figure~\ref{fig:scale_num_client_lora_count}).

\textit{Insight 1: in our disaggregated design, cache capacity is the primary scaling bottleneck when serving a large number of LLM instances, which motivates scaling the LoRA Server itself.}

\subsubsection{Scale with Server Resources}
\label{sec:exp_scale_server}

\begin{figure}[tbp]
    \vspace{-4pt}
    \centering
    \begin{subfigure}{0.95\linewidth}
        \centering
        \includegraphics[width=\linewidth, height=2.1cm]{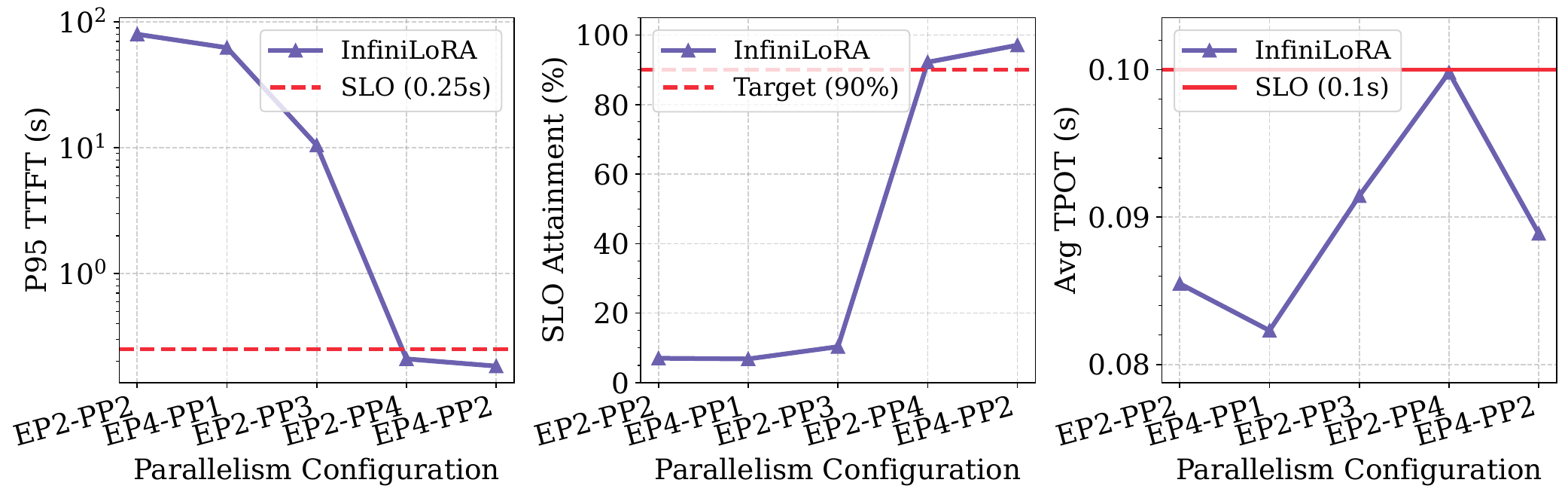}
        \vspace{-13pt}
        \caption{P95 TTFT, SLO attainment rate and average TPOT from left to right using different LoRA Server parallelism configuration.}
        \label{fig:scale_num_server}
    \end{subfigure}
    \begin{subfigure}{0.95\linewidth}
        \centering
        \includegraphics[width=\linewidth, height=3.4cm]{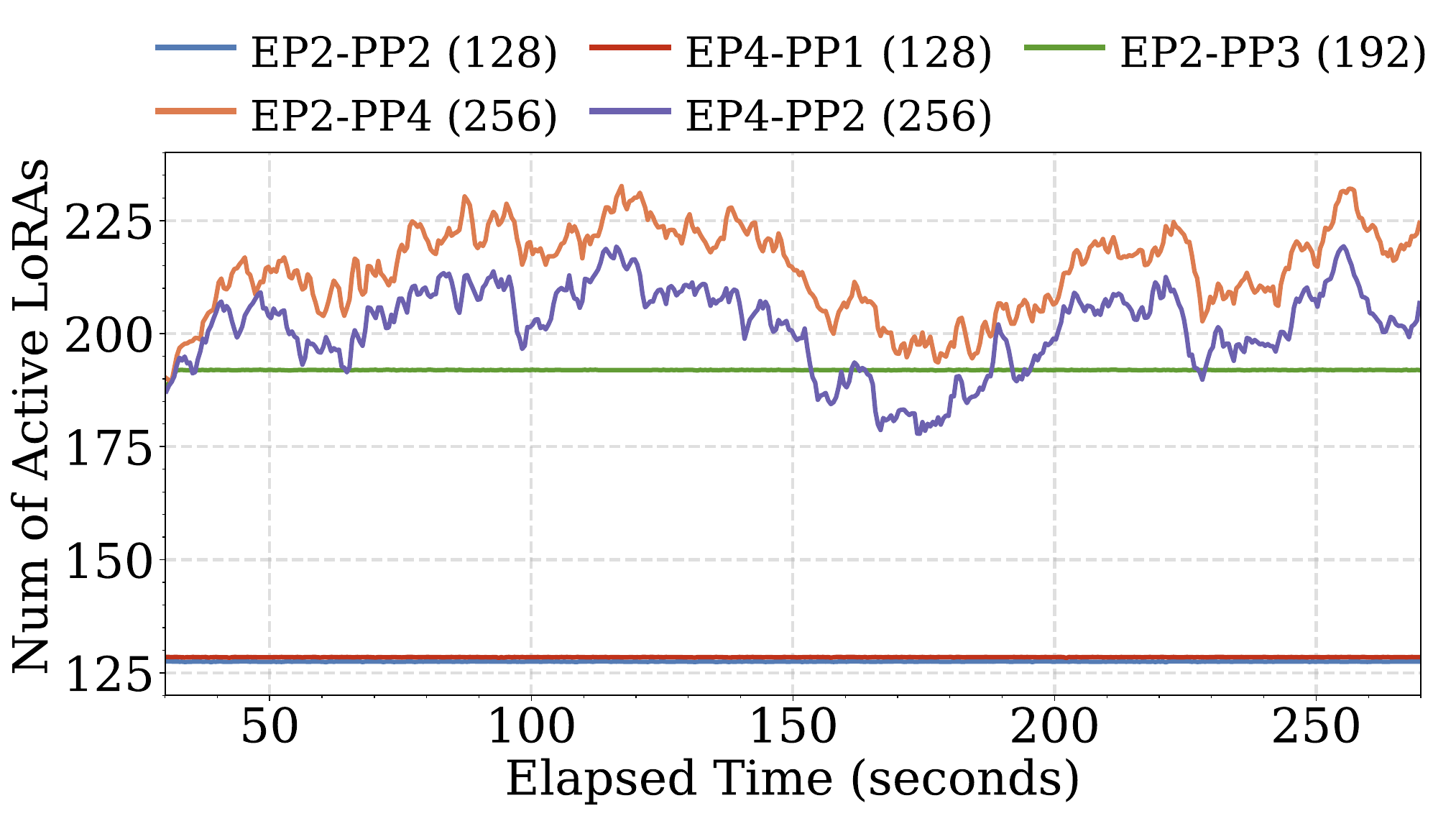}
        \vspace{-13pt}
        \caption{The evolution of active LoRA counts over times. Numbers in parentheses denote the server's LoRA cache capacity.}
        \label{fig:scale_num_server_lora_count}
    \end{subfigure}
        \vspace{-10pt}
    \caption{Performance of scaling LoRA Server resources (we keep resources for LLM instances unchanged, using Qwen3-30B-A3B model and request rate=35 req/s).}
    \Description{Need to modify, just a simplified diagram}
    \label{fig:scale_num_server_2fig}
    \vspace{-9pt}
\end{figure}

With four LLM instances (each serving one Qwen3-30B-A3B model) and a fixed request rate of 35 req/s, we scale the LoRA Server by provisioning 4, 6, and 8 GPUs using five different parallelism configurations. 
As shown in Figure~\ref{fig:scale_num_server}, increasing LoRA Server resources significantly improves P95 TTFT and SLO attainment by expanding cache capacity. 
\chen{\chen{For LoRA cache capacities of 128, 192, and 256, the model in Section~\ref{sec:resource_provision} predicts IARs of 83.0\%, 92.2\%, and 100.0\%, respectively. The observed TTFT trend qualitatively supports the model's prediction that larger caches alleviate adapter-related admission pressure.}}
Under the 8-GPU configuration, a hybrid layout biased toward larger expert parallelism (i.e., $EP_{4}\text{-}PP_{2}$) achieves lower TPOT, as intra-node synchronization overhead is small while more server GPUs are used to process each layer, aligning with our analysis in Section~\ref{sec:lora_execution}. Figure~\ref{fig:scale_num_server_lora_count} shows the number of concurrently active adapters. For the 4-GPU and 6-GPU configurations, limited cache capacity caps the number of active adapters, directly leading to degraded tail TTFT and lower SLO attainment observed in Figure~\ref{fig:scale_num_server}.

\textit{Insight 2: provisioning additional GPUs for LoRA Server effectively alleviates the cache capacity bottleneck, and that the choice of parallelism strategy further influences the efficiency of the disaggregated architecture.}

\begin{table}[tbp]
\centering
\caption{Latency breakdown of LoRA execution and base MoE computation under varying parallelism (\textmu s).}
\vspace{-8pt}
\resizebox{\columnwidth}{!}{%
\begin{tabular}{c|cccc|cccc}
\hline
\multirow{2}{*}{$EP_{x}\text{-}PP_{y}$} & \multicolumn{4}{c|}{Batch Size = 128} & \multicolumn{4}{c}{Batch Size = 256} \\
\cline{2-9}
 & Recv & LoRA & Send & MoE & Recv & LoRA & Send & MoE \\
\hline
$EP_1\text{-}PP_8$ & 243 & 342 & 384 & 493 & 527 & 526 & 734 & 762 \\
$EP_2\text{-}PP_4$ & 155 & 212 & 221 & 493 & 279 & 315 & 402 & 764 \\
$EP_4\text{-}PP_2$ & 145 & 142 & 207 & 492 & 246 & 204 & 396 & 763 \\
$EP_8\text{-}PP_1$ & 173 & 163 & 255 & 493 & 310 & 205 & 441 & 763 \\
\hline
\end{tabular}%
}
\label{tab:time_breakdown_on_epx_ppy}
\vspace{-0.1cm}
\end{table}

\subsection{Ablation study}

We conduct an ablation study to quantify the impact of each optimization. Experiments use the Mixtral model with a fixed request rate of 25 req/s and 256 adapters in total. InfiniLoRA allocates 2 GPUs to the LoRA Server (cache capacity of 104) and 6 GPUs to three LLM instances, while S-LoRA uses all 8 GPUs to run four LLM instances (total cache capacity of 100); all other settings match the end-to-end evaluation.
We start from a disaggregated baseline that separates LoRA adapters from the base model (\texttt{+disagg}), then add communication–computation overlap (\texttt{+overlap}), followed by layer-wise adapter loading (\texttt{+loading}). The full InfiniLoRA system further incorporates hardware-specialized kernels (\texttt{+kernel}).

As shown in the results, despite having a slightly larger cache capacity (104 over 100), \texttt{+disagg} alone increases tail TTFT from 0.78~s to 0.99~s, indicating that a naive disaggregated architecture actually degrades performance due to the additional communication overhead it introduces. By incrementally adding the remaining optimizations, InfiniLoRA reduces P95 TTFT by $11\times$, lowers average TPOT by 30\%, and achieves a 100\% SLO attainment rate, highlighting the complementary benefits of each technique.

\textit{Insight 3: disaggregation alone is insufficient—its benefits are realized with critical-path optimization.}

\begin{figure}[tbp]
    \vspace{-4pt}
    \centering
    \includegraphics[width=0.95\columnwidth, height=2.2cm]{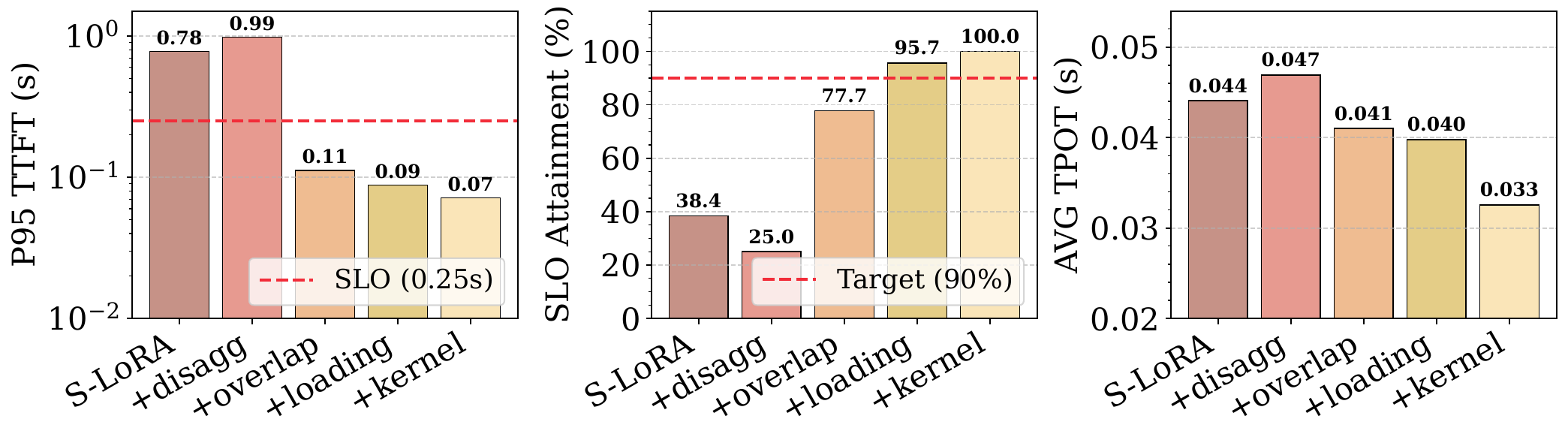}
        \vspace{-10pt}
    \caption{Ablation study for quantifying the effectiveness of individual optimization techniques. \texttt{+kernel} represents the fully optimized system with all techniques enabled.}
    \Description{Need to modify, just a simplified diagram}
    \label{fig:ablation_study}
    \vspace{-0.3cm}
    \vspace{-2pt}
\end{figure}

\subsection{Microbenmark}

\subsubsection{LoRA data layout}
\label{sec:apd:microb_data_layout}

\begin{figure}[tbp]
    \centering
    \includegraphics[width=0.95\columnwidth]{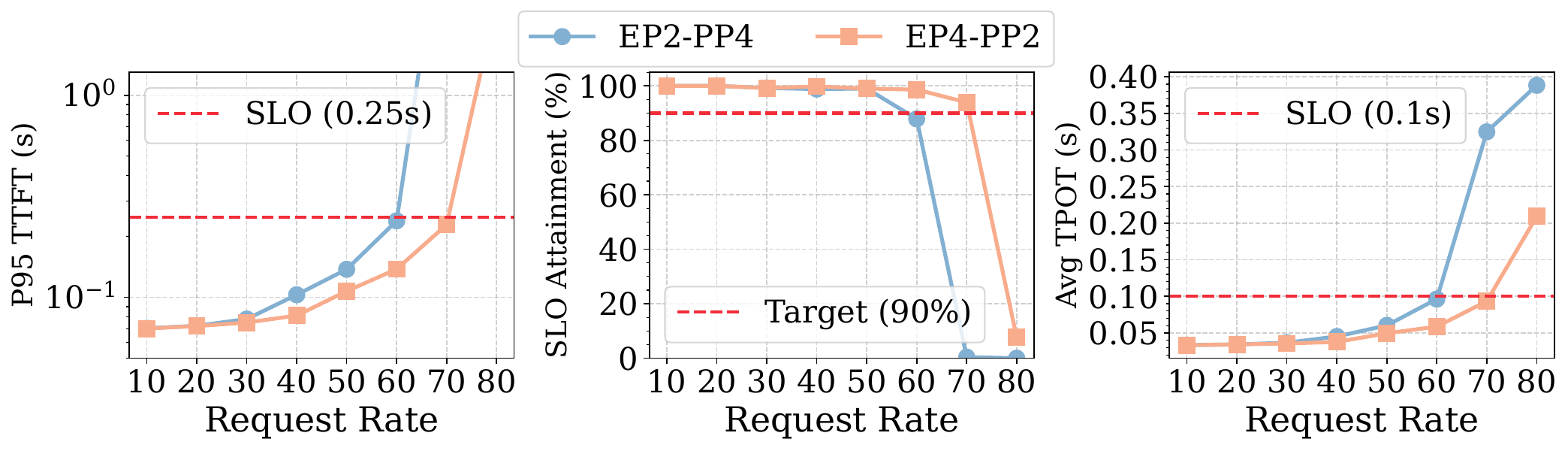}
        \vspace{-10pt}
    \caption{End-to-end performance comparison of two LoRA data layouts ($EP_2\text{-}PP_4$ and $EP_4\text{-}PP_2$) under the same LoRA cache capacity.}
    \Description{Need to modify, just a simplified diagram}
    \label{fig:micro_data_layout}
    \vspace{-0.3cm}
\end{figure}

Table~\ref{tab:time_breakdown_on_epx_ppy} compares four EP--PP layouts on an 8-GPU LoRA Server using Mixtral-8x7B workloads at batch sizes 128 and 256. Pure pipeline parallelism ($EP_1\text{-}PP_8$) incurs high execution latency even under ideal stage balancing, while pure expert parallelism ($EP_8\text{-}PP_1$) has higher computation and communication latency than $EP_4\text{-}PP_2$. Among hybrid layouts, $EP_4\text{-}PP_2$ provides greater per-stage processing capacity and is less sensitive to pipeline imbalance than $EP_2\text{-}PP_4$. Figure~\ref{fig:micro_data_layout} confirms its better end-to-end performance at equal cache capacity.

\subsubsection{LoRA kernels}
\label{sec:apd:microb_lora_kernels}

\begin{figure}[tbp]
    \centering
    \includegraphics[width=0.68\columnwidth]{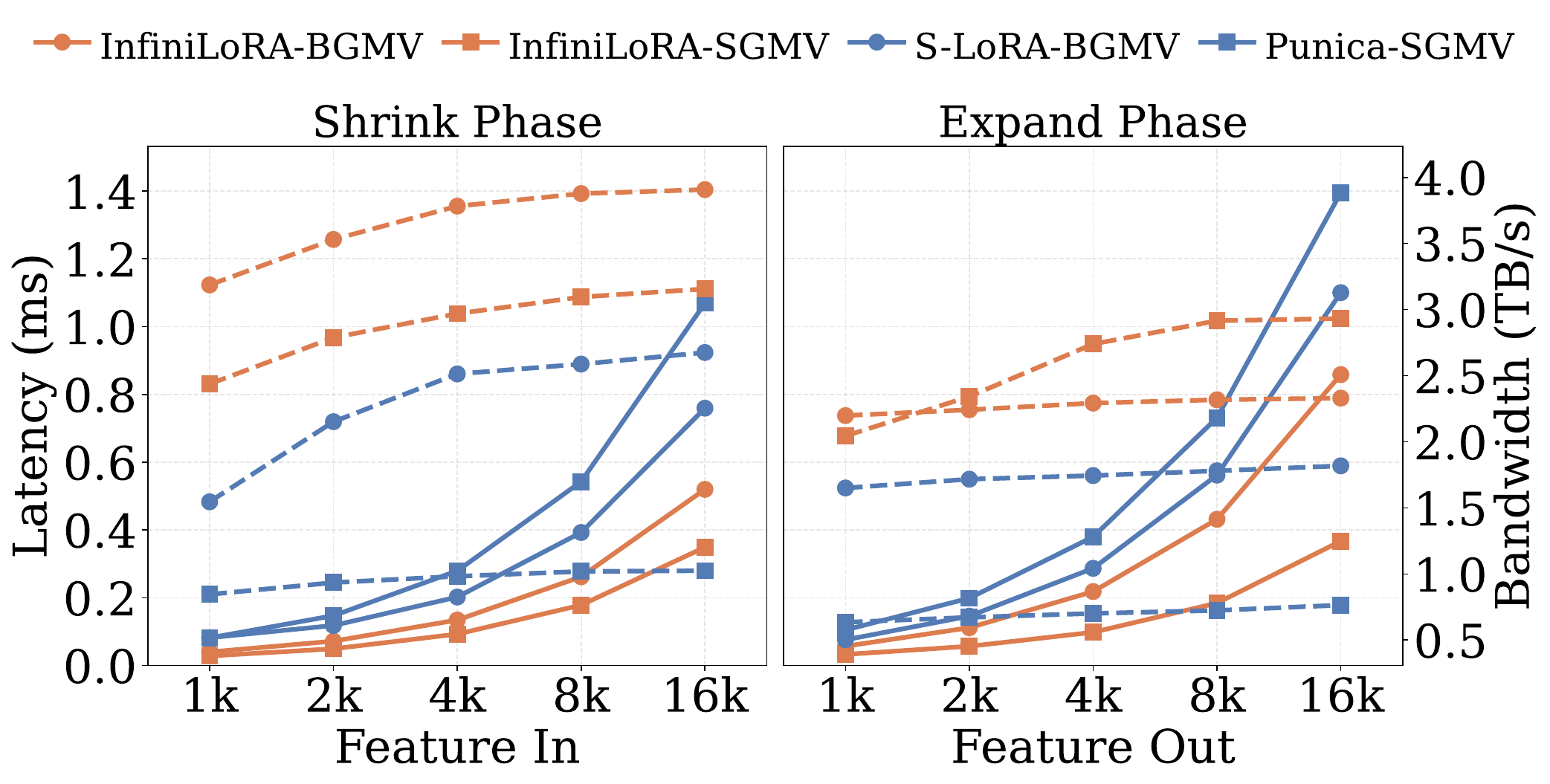}
        \vspace{-10pt}
    \caption{Latency and bandwidth characterization of different LoRA kernels across shrink/expand phases. Dashed lines denote bandwidth and solid lines denote latency.}
    \Description{Need to modify, just a simplified diagram}
    \label{fig:lora_kernel}
    \vspace{-0.3cm}
\end{figure}

Figure~\ref{fig:lora_kernel} compares our kernels with Punica~\cite{chen2023punicamultitenantloraserving} and S-LoRA~\cite{sheng2023slora} using 512 adapters, batch size 1024, and Zipf-distributed invocations ($s=1.2$). InfiniLoRA improves latency and bandwidth utilization over both baselines. InfiniLoRA-BGMV performs well in shrink, but larger writeback volumes slow expand. By grouping tokens sharing an adapter into a single GEMM, InfiniLoRA-SGMV reduces memory traffic and maintains lower latency.

\section{Related works}

\noindent\textbf{Disaggregation design.}
Prefill-decode disaggregation frameworks~\cite{Pratyush2024splitwise, Zhong2024distserve, Hu2025ShuffleInfer, qin2025mooncake, hu2024memservecontextcachingdisaggregated} decouple the compute-intensive prefill phase from the memory-intensive decode phase to enable phase-specific resource management. Recent works~\cite{zhu2025megascleinfer, stepfun2025step3largeaffordablemodelsystem, xiao2026huaweicloudmodelasaservicecloudmatrix384, liu2025expertasaserviceefficientscalablerobust} further explore finer-grained Attention-FFN disaggregation, enabling independent scaling, tailored parallelism, and heterogeneous deployment for each module. 

\noindent\textbf{Request batching.} 
Recent works like Punica~\cite{chen2023punicamultitenantloraserving} and S-LoRA~\cite{sheng2023slora} propose to batch requests with heterogeneous LoRA adapters through on-demand loading, using optimized CUDA kernels and unified memory management to efficiently serve concurrent adapters. dLoRA~\cite{dLoRA24osdi} introduces dynamic switching between merged and unmerged LoRA inference modes, and co-migrates requests and adapters across worker replicas to adapt to varying workload distributions.

\noindent\textbf{Request scheduling strategy.} 
\chen{Chameleon~\cite{Iliakopoulou2025Chameleon} improves serving quality within available local GPU LoRA capacity through cache management and adapter-aware scheduling. 
Cannikin~\cite{zhu2025Cannikin} optimizes SLO attainment for the worst-served adapter service. 
InfiniLoRA enables independent adapter-resource provisioning: sufficient server cache can avoid capacity induced degradation, at the cost of activation transfers and client--server coordination that local execution avoids. 
InfiniLoRA prescribes no request-priority policy, so integrating these strategies with its expanded cache is possible.}

\noindent\textbf{LoRA cache management.} 
Toppings~\cite{li2025toppings} leverages CPU cores for LoRA computation in prefill to address cache-miss latency, but limited host memory bandwidth makes it unsuitable for low-latency decoding.
FASTLIBRA~\cite{zhang2025improvingservingperformancemultilora} jointly manages LoRA adapters and KV caches within a unified HBM pool to reduce TTFT through dependency-aware eviction and swapping.
LoRAServe~\cite{jaiswal2025servingheterogeneousloraadapters} optimize LoRA placement among LLM instance clusters by dynamically rebalancing adapters across GPUs. Despite LoRAServe's nominal resemblance to our approach, its targets is fundamentally different.

\chen{Existing coupled systems mitigate capacity-related cache misses and loading overhead through cache management, request scheduling, and adapter placement. 
But their adapter cache capacity remains tied to the available GPU memory of LLM instances.
InfiniLoRA instead pools adapter storage and computation on dedicated server GPUs, allowing cache expansion without additional base-model copies, and further introduces optimizations on top of this decoupled design.}
\section{Conclusion}

We introduce InfiniLoRA, a disaggregated serving system that decouples LoRA execution from base model inference to resolve scalability bottlenecks in multi-tenant scenarios. By leveraging a shared LoRA Server with parallelism-aware execution, SLO-driven provisioning, and critical-path optimizations, InfiniLoRA can flexibly scale LoRA cache capacity without interfering with the LLM inference pipeline. Our evaluation demonstrates that InfiniLoRA serves a $3.05\times$ higher average request rate while satisfying latency SLOs, and improves the percentage of LoRA adapters satisfying the SLO requirement by 54.0\% compared to existing systems.

\begin{acks}

We thank the anonymous reviewers for their constructive feedback.
This work was supported by Fundamental and Interdisciplinary Disciplines Breakthrough Plan of the Ministry of Education of China (JYB2025XDXM113).
\end{acks}

\clearpage
\bibliographystyle{ACM-Reference-Format}
\bibliography{sample-base}

\clearpage

\end{document}